\documentclass[journal, twocolumn, 10pt]{IEEEtran}

\usepackage[T1]{fontenc}
\usepackage{lipsum}
\usepackage{blindtext}
\usepackage{multicol}
\usepackage{graphicx}
\usepackage{cite}
\usepackage{amssymb,amsmath}
\usepackage{algorithm}
\usepackage{algpseudocode}
\usepackage{flushend}
\usepackage{multirow}
\usepackage{hhline}
\usepackage{tcolorbox}

\usepackage{graphicx}
\usepackage{cite}
\usepackage{amssymb,amsmath}
\usepackage{algorithm}
\usepackage{color,soul,colortbl}
\usepackage{cleveref}
\usepackage{subcaption}
\usepackage{booktabs}
\usepackage{xcolor}
\usepackage{caption}
\usepackage{multirow}
\usepackage{hhline}
\usepackage{psfrag}

\usepackage{widetext}
\usepackage{flushend}
\usepackage{cuted}
\usepackage{float}
\usepackage{lipsum}

\newcounter{tempEquationCounter}
\newcounter{thisEquationNumber}
\newenvironment{floatEq}
{\setcounter{thisEquationNumber}{\value{equation}}\addtocounter{equation}{1}
\begin{figure*}[!t]
\normalsize\setcounter{tempEquationCounter}{\value{equation}}
\setcounter{equation}{\value{thisEquationNumber}}
}
{\setcounter{equation}{\value{tempEquationCounter}}
\hrulefill\vspace*{4pt}
\end{figure*}
}

\usepackage{amsthm}

\newtheorem{lemma}{Lemma}

\newtheorem{remark}{Remark}

\newtheorem{definitioninner}{Definition}

\newenvironment{definition}[1]{%
  \renewcommand\thedefinitioninner{#1}
  \definitioninner
}{\enddefinitioninner}

\theoremstyle{plain}

\theoremstyle{plain}

\theoremstyle{plain}

\providecommand{\lemmaname}{Lemma}
\providecommand{\propositionname}{Proposition}
\providecommand{\theoremname}{Theorem}
\providecommand{\lemmaname}{Lemma}
\providecommand{\propositionname}{Proposition}
\providecommand{\theoremname}{Theorem}
\makeatother
\providecommand{\lemmaname}{Lemma}
\providecommand{\propositionname}{Proposition}
\providecommand{\theoremname}{Theorem}

\definecolor{G}{gray}{0.9}
\definecolor{LC}{rgb}{0.88,1,1}

\newcommand{\V}[1]{\mathbf{\lowercase{#1}}}                  
\newcommand{\M}[1]{\mathbf{\uppercase{#1}}}                  
\newcommand{\RM}[1]{{\mathrm{#1}}}  
 
\usepackage[nolist]{acronym}

\begin{document}
\title{{\LARGE{When Probabilistic Shaping Realizes Improper Signaling for Hardware Distortion Mitigation}}}
\author{{ Sidrah Javed,~\IEEEmembership{Student~Member,~IEEE,} Ahmed Elzanaty,~\IEEEmembership{Member,~IEEE,}  Osama Amin, ~\IEEEmembership{Senior~Member,~IEEE,} \\ Basem Shihada, ~\IEEEmembership{Senior~Member,~IEEE,}  and Mohamed-Slim Alouini,~\IEEEmembership{Fellow,~IEEE}} 
\thanks{S. Javed, A. Elzanaty, O. Amin, B.Shihada and M.-S. Alouini are with CEMSE Division,  King Abdullah University of Science and Technology (KAUST),  Thuwal, Makkah Province, Saudi Arabia.   E-mail: \{sidrah.javed, ahmed.elzanaty, osama.amin, basem.shihada, slim.alouini\}@kaust.edu.sa } }
 \maketitle
 \begin{acronym}
\acro{5G}{fifth generation}
\acro{BER}{bit error rate}
\acro{AWGN}{additive white Gaussian noise}
\acro{CDF}{cumulative distribution function}
\acro{CSI}{channel state information}
\acro{FDR}{full-duplex relaying}
\acro{HDR}{half-duplex relaying}
\acro{IC}{interference channel}
\acro{IGS}{improper Gaussian signaling}
\acro{MHDF}{multi-hop decode-and-forward}
\acro{SIMO}{single-input multiple-output}
\acro{MIMO}{multiple-input multiple-output}
\acro{MISO}{multiple-input single-output}
\acro{MRC}{maximum ratio combining}
\acro{PDF}{probability density function}
\acro{PGS}{proper Gaussian signaling}
\acro{RSI}{residual self-interference}
\acro{RV}{random vector}
\acro{r.v.}{random variable}
\acro{HWD}{hardware distortion}
\acro{cHWD}[HWD]{Hardware distortion}
\acro{AS}{asymmetric signaling}
\acro{GS}{geometric shaping}
\acro{PS}{probabilistic shaping}
\acro{HS}{hybrid shaping}
\acro{SISO}{single-input single-output}
\acro{QAM}{quadrature amplitude modulation}
\acro{PAM}{pulse amplitude modulation}
\acro{PSK}{phase shift keying}
\acro{DoF}{degrees of freedom}
\acro{ML}{maximum likelihood}
\acro{MAP}{maximum a posterior}
\acro{SNR}{signal-to-noise ratio}
\acro{SCP}{successive convex programming}
\acro{RF}{radio frequency}
\acro{CEMSE}{Computer, Electrical, and Mathematical Sciences and Engineering}
\acro{KAUST}{King Abdullah University of Science and Technology}
\acro{DM}{distribution matching}
\end{acronym}

\begin{abstract}
\acp{cHWD} render drastic effects on the performance of communication systems. They are recently proven to bear asymmetric signatures; and hence can be efficiently mitigated using \acs{IGS}, thanks to its additional design degrees of freedom. Discrete \acs{AS}  can practically realize the \acs{IGS} by shaping the signals' geometry or probability. In this paper, we adopt the \acs{PS} instead of uniform symbols to mitigate the impact of \acp{cHWD} and derive the optimal maximum a posterior detector. Then, we design the symbols' probabilities to minimize the error rate performance while accommodating the improper nature of \acs{cHWD}. Although the design problem is a non-convex optimization problem, we simplified it using successive convex programming and propose an iterative algorithm. We further present a \acs{HS} design to gain the combined benefits of both \acs{PS} and \acs{GS}. Finally, extensive numerical results and Monte-Carlo simulations highlight the superiority of the proposed \acs{PS} over conventional uniform constellation and  \acs{GS}. Both \acs{PS} and \acs{HS} achieve substantial improvements over the traditional uniform constellation and \acs{GS} with up to one order magnitude in error probability and throughput. 

\end{abstract}
\acresetall
\section{Introduction} 
Exponentially rising demands of high data rates and reliable communications given the limited power and bandwidth resources impose enormous challenges on the next generation of wireless communication systems  \cite{chettri2019comprehensive,javed2020journey}. Various research contributions propose new configurations and novel techniques to address these challenges \cite{razavi1997design, abidi1995direct}. Nonetheless, the performance of such systems can be highly degraded by the hardware imperfections in \ac{RF} transceivers \cite{buzzi2016survey,schenk2008rf,krishnan2015impact}. Such imperfections give rise to additive signal distortions emerging from the phase noise, 
mismatched local oscillator, 
imperfect high power amplifier/low noise amplifier, non-linear 
amplitude-to-amplitude and amplitude-to-phase transfer  \cite{xia2015hardware,bjornson2013new, matthaiou2013two,bjornson2014massive,duy2015proactive,
bjornson2013capacity,studer2010mimo}.  Various contributions emphasized the distinct improper behavior of these \acp{HWD} \cite{javed2017on, soleymani2019improper,javed2018improper,alsmadi2018ssk}, which requires effective compensation techniques to meet the performance demands. 

\subsection{Motivation}
The \ac{IGS} is proven as an effective scheme to mitigate the deteriorating effects due to the existence of improper noise or interference in wireless communication systems. More precisely,  
\ac{IGS} is a generalized complex signaling that allows the signal components to be  correlated and/or to have unequal power, as opposed to  proper Gaussian signaling\cite{neeser1993proper}. \ac{IGS} offers an additional 
\ac{DoF} in signaling design characterized by the circularity coefficient \cite{ollila2008circularity}. 
Several studies highlight the significance of \ac{IGS} to improve the system performance under improper interference \cite{lameiro2015benefits,amin2017overlay,hedhly2017interweave, lagen2014decentralized,kurniawan2015improper, mahady2019sum,lin2018multi,lameiro2018improper}. Recent studies quantified the impact of \ac{IGS} in dampening 
improper noise effects in multi-antenna or multi-nodal system settings \cite{javed2016impact,javed2017asymmetric,javed2018multiple,kim2012asymmetric,javed2018improper,gaafar2016alternate},
\ac{IGS} has emerged as promising candidate to improve the  average achievable rate performance in multi-antenna systems suffering from  \ac{HWD} \cite{javed2017asymmetric,javed2018multiple}. 
Moreover, \ac{IGS} benefits can also be reaped in various full-duplex/half-duplex relay settings by effectively compensating the residual self-interference, inter-relay interference and/or  \ac{HWD} \cite{kim2012asymmetric,javed2018improper,gaafar2016alternate}.
Additionally, the ergodic rate maximization and outage probability minimization based on a generalized error model for hardware impairments in \ac{SIMO} and \ac{MIMO} systems is studied in \cite{javed2018multiple,javed2017asymmetric}.

\subsection{Background}
Despite the overwhelming benefits of \ac{IGS}, it is
practically infeasible owing to the high detection complexity and unbounded peak-to-average power ratio \cite{santamaria2018information,javed2020journey}.
This motivated the researchers to design some equivalent finite and discrete \ac{AS} schemes for practical implementation. Improper discrete constellation, or AS,  entails redesigning the symmetric discrete signal constellation to convert it into an asymmetric signal \cite{javed2020journey}. Several studies focused on \ac{GS} as a possible designing scheme to improve system performance. \ac{GS} transforms equally spaced symbols to unequally spaced symbols (due to correlated and/or unequal power distribution
between quadrature components of the symbols) in a distinct geometric envelop such as ellipse \cite{lopez2019design}, parallelogram \cite{javed2019asymmetric,santamaria2018information} or some irregular envelop \cite{foschini1974optimization}. A family of improper discrete constellations generated by widely linear processing of a square $M$-ary \ac{QAM} depict parallelogram envelop \cite{santamaria2018information}. Similarly, \ac{GS} based on optimal translation and rotation also yields parallelogram envelop \cite{javed2019asymmetric}. However, conditioned on high \ac{SNR} and higher order QAM, the optimal constellation is the intersection of the hexagonal lattice/packing with an ellipse where the eccentricity determines the circularity coefficient \cite{lopez2019design}. \ac{GS} has emerged as a competent player to reduce shaping loss and improve reception at lower signal-to-noise ratios in terrestrial broadcast systems \cite{loghin2016non,zollner2013optimization}. \ac{GS} parameters can be designed for diverse objectives such as capacity maximization \cite{santamaria2018information}, \ac{BER} reduction \cite{javed2019asymmetric}, and symbol error probability minimization \cite{lopez2019design}.
Although the asymmetric discrete family of constellations is practical, they exhibit two types of loss, i.e., shaping loss and packing loss in approaching \ac{IGS} theoretical limits \cite{santamaria2018information}. 

\subsection{Related Work}
Most of the efforts to close the gap between AS and ideal \ac{IGS} are concentrated around \ac{GS} with a limited focus on \ac{PS} as another way to 
implement AS for \ac{HWD}. Given a fixed number of
symbols and the symbol locations, an asymmetric constellation can be obtained by adjusting the symbol probabilities \cite{thaiupathump2000asymmetric}.   \ac{PS}  maps equally distributed input bits into constellation symbols with non-uniform prior probabilities \cite{batshon2017coded}. This can be achieved using \ac{DM}  for rate adaptation such as constant composition \ac{DM}  \cite{SchBoc:16},  adaptive arithmetic \ac{DM}  \cite{schulte2014zero}, syndrome \ac{DM}  \cite{Cirinothesis:18,ElzGioChisyndrome:19} \ac{DM}-based compressed sensing  \cite{dia2018compressed,ElzGioChi:19}.
 
\ac{PS}-based schemes have been employed to enhance the system performance in optical fiber communications (OFC) and free-space optics (FSO).  In OFC, multiple transformations are presented to approach Gaussian channel capacity using \ac{PS} including prefix codes \cite{kschischang1993optimal,forney1984efficient}, many-to-one mappings combined with a turbo code \cite{yankov2014constellation}, distribution matching \cite{buchali2016rate} and cut-and-paste method \cite{cho2016low}.  Furthermore, multidimensional coded modulation format with hybrid probabilistic and geometric constellation shaping can effectively compensate non-linearity and approach Shannon limits in OFC \cite{CaiBat:17}. Coded modulation scheme with \ac{PS} aims to solve the shaping gap and coarse mode granularity problems \cite{bocherer2015bandwidth}.
Interested reader can read the classic work \cite{dunbridge1967asymmetric} for the 
design guidelines of \ac{AS}  in the
coherent Gaussian channel with equal signal energies and unequal a priori probabilities. Probabilistic amplitude shaping is another concept that can only be used for symmetric constellation with coherent modulation, which greatly limits its application \cite{pan2016probabilistic}. For FSO, a practical and capacity achieving \ac{PS}  scheme with adaptive coding modulation is proposed with intensity modulation/direct detection~\cite{elzanaty2020adaptive}. 

 The concept of \ac{PS} is widely employed in the OFC and FSO systems. However, it is quite not well investigated in wireless communication systems and only a few studies have contributed in this domain  \cite{gultekin2017constellation,iimori2017constellation}. For example, enumerative amplitude shaping is proposed as a constellation shaping scheme for IEEE 802.11 which renders Gaussian distribution on the constituent constellation \cite{gultekin2017constellation}. Moreover, \ac{PS} has been proposed to maximize the mutual information between transmit and receive signals for non-linear distortion effects in \ac{AWGN} channels \cite{iimori2017constellation}.  To the best of authors' knowledge, \ac{PS} has not been used to enhance the error performance or to realize the  \ac{IGS} for wireless communication systems with \ac{HWD}.  

\subsection{Contributions}
In this paper, we propose \ac{PS} as a method to realize improper signaling, which is beneficial in mitigating the impact of HWD on the BER performance. Motivated by \ac{IGS}'s theoretical results in various scenarios \cite{javed2020journey} and the issues associated with \ac{GS}, such as high shaping gap and coarse granularity, we adopt PS to realize the  \ac{IGS} scheme and combat \ac{HWD} to assure reliable communications. 
In the following, we summarize the main contributions as:
\begin{itemize}
\item We derive the optimal \ac{MAP}  detector for a discrete \acs{AS} and carry out \ac{BER}  analysis for the adopted  \ac{HWD} communication system. 
\item We design the probabilistic shaped \ac{AS}  under power and rate constraints for hardware distorted system and propose adaptive algorithm that tune the  symbol probabilities for \ac{PS} to minimize the \ac{BER} performance. 
\item We further suggest a hybrid shaped \ac{AS}  scheme that reaps benefits of both \ac{PS} and \ac{GS} and present an adaptive algorithm that tune both signal probability and  shaping parameters.   

\item Finally, we present numerical Monte-Carlo simulations to validate the performance of the proposed techniques and compare the  \ac{BER}  and throughput performance of PS, GS, and \ac{HS}  in AWGN and Rayleigh fading channels.
\end{itemize}
\subsection{Paper Organization and Notation}
The rest of the paper is organized as: Section II describes statistical signal characteristics,  \ac{HWD} model, and optimal receiver for the adopted  \ac{HWD} system. 
 In section III, we present the error probability analysis using the union bound on pairwise error probability and derive instantaneous \ac{BER}  for generalized $M$-ary modulation scheme. Next, we propose \ac{PS}  design using \ac{SCP}  algorithm and some toy examples for comprehensive illustration in section IV. Later, \ac{HS}  parameterization and design along with the respective \ac{MAP}  and error probability analysis is carried out in section V, followed by the numerical results in Section VI and the conclusion in Section~VII. 

\textit{Notations:} In this paper, $|a|$ and $a^*$ represent the absolute and complex conjugate of a scalar complex number $a$. The probability of an event $A$ is defined as ${\rm Pr}(A)$. The notations $f_z(z)$ and $f_{z|y}(z|y)$ denote the \ac{PDF}  and conditional PDF of a random variable (r.v.) $z$ given $y$. The operator ${\mathbb{E}}[.]$ denotes the expected value. Considering a r.v. $\Lambda$, the real/in-phase and imaginary/quadrature-phase components of $\Lambda$ are denoted as $\Lambda_{\rm{I}}$ and $\Lambda_{\rm{Q}}$, respectively.
Moreover, $f'(x)$ denotes the first order derivative of $f(x)$ with respect to $x$. Additionally, $\mathcal{Z}^{+}$ represents a set of positive integers. $\V{v} = [ v_I  \quad v_Q  ]^{\rm T}$ is the real-composite vector representation of the complex number $v = v_I + i\, v_Q$. Furthermore, $x^{(k)}$ and $\mathbf{p}^{(k)}$ represent the instance values of the  variable $x$ and  vector $\mathbf{p}$, respectively, in the $k^{\rm th}$ iteration of an algorithm.

\section{System Description}\label{SecII}
Impropriety incorporation is crucial for the systems dealing with improper signals, noise, or interference.  Such characterization helps in  meticulous system modeling, accurate performance analysis, and optimum signaling design. We begin by presenting the statistical signal model to introduce some preliminaries of the impropriety characterization. This will help to comprehend the impropriety concepts in the adopted system model with \ac{HWD}. Then, the transceiver \ac{HWD} model is described, and the optimal receiver is  derived. 

\subsection{Statistical Signal Model}
\label{ssec:SignalModel}
The impropriety characterization of a random variable (r.v.) $x$ involves the identification and extent of improperness described by the pseudo-variance and circularity coefficient, respectively.

 \begin{definition} {1}
The pseudo-variance  of $x$ is defined as $\tilde \sigma_{x}^2 ={\mathbb{E}}[x^2]$ as opposed to the conventional variance $\sigma_{x}^2 = {\mathbb{E}}[|x|^2]$ \cite{neeser1993proper}. A null $\tilde \sigma_{x}^2$ signifies a proper complex r.v. whereas a non-zero $\tilde \sigma_{x}^2$ identifies an improper complex r.v.
 \end{definition}
 \begin{definition} {2}
 The degree of improperness is given by the circularity coefficient  $\mathcal{C}_x \triangleq {\left| \tilde{\sigma}_{x}^2 \right|} / {\sigma_{x}^2},$ where $0\leq \mathcal{C}_x \leq 1$ \cite{ollila2008circularity}. $\mathcal{C}_x = 0$ indicates proper or symmetric signal and $\mathcal{C}_x = 1$ indicates maximally improper or maximally asymmetric signal. 
 \end{definition}
  Evidently, the pseudo-variance is bounded, i.e., ${0 \le \left| \tilde{\sigma}_{x}^2 \right|  \le {\sigma}_{x}^2 }$. Interestingly, a complex Gaussian random variable $v=a+ib$ can be fully described as,    $v \sim \mathcal{CN}(m_v , \sigma_{v}^2 , \tilde{\sigma}_{v}^2 )$, where $m_v$ is the statistical mean of $v$ \cite{javed2020journey}. The PDF of $v$ with augmented representation 
   $\underline{v} = [{v} \;  {v}^{*}]^{\rm {T}}$ is given as \cite{van1995multivariate}
\begin{equation} \label{eqpdf}
p\left( {v} \right)\! = \!\frac{1}{{\sqrt {{\pi ^{2}}\left| {{\M{R}_{ \underline{v}  \underline{v} }}} \right|} }}\exp \left\{ {\! - \frac{1}{2}\left( { \underline{v}\!  -\! {\V{\mu} _{ \underline{v} }}} \right)^{\RM{H}} \!\M{R}_{ \underline{v}  \underline{v} }^{ - 1}\!\left( { \underline{v}\!  - \!{\V{\mu} _{ \underline{v} }}} \right)} \right\},
\end{equation} 
  where ${\mu}_{\underline{v}}$ and
  $\M{R}_{\underline{vv}}$ are the mean and augmented covariance matrix of $\underline{v}$, respectively
 \cite{picinbono1996second}, i.e.,
 \begin{equation}
{\mu}_{\underline{v}} =  \left[{\begin{array}{*{20}{c}}
{\mathbb{E}}[a]+i{\mathbb{E}}[b]\\
{\mathbb{E}}[a]-i{\mathbb{E}}[b]\end{array}} \right], \; {\M{R}_{\underline{vv}}} = \left[ {\begin{array}{*{20}{c}}
\sigma_{v}^2 &{\tilde{\sigma}_{v}^2}\\
{\tilde{\sigma}_{v}^{2*}}&\sigma_{v}^2
\end{array}} \right]. 
\end{equation}

\subsection{Transceiver Hardware Distortion Model}
\label{ssec:DistModel}
Consider a single-link wireless communication system suffering from various hardware impairments. The non-linear transfer functions of various transmitter \ac{RF} stages, such as digital-to-analog converter, band-pass filter and  high power amplifier result in accumulative additive distortion noise ${\eta_{\mathrm{t}}}\sim{\mathcal{CN}}(0,\kappa _{\mathrm{t}},\tilde \kappa_{\mathrm{t}} )$, where  $| \tilde{\kappa}_{\rm t} | \leq \kappa_{\rm t}$  \cite{bjornson2013new,schenk2008rf}. 
These distortions raise the noise floor of the transmitted signal ${x_\mathrm{tx}} = {x_{m}} + {\eta_{\rm{t}}} $, where $x_m$ is the single-carrier band-pass modulated signal taken from $M$-ary QAM, $M$-ary \ac{PSK}, or $M$-ary \ac{PAM} constellation with a probability mass function $p_{m} \triangleq p_X(x_m)$ rendering the transmission probability of symbol $x_m$, and ${\mathbf p} \triangleq [p_{1},p_{2},\cdots,p_{M}]$.
Let us define the set that includes all possible symbol distributions as
\begin{align}
\mathbb{S}=\Bigg\{& {\bf p} :       {\bf p} = [p_{1},p_{2},\cdots, p_{M}] , \nonumber \\
&\sum_{j=1}^{M} {p}_j=1,    p_{j}\geq 0, \,\forall j \in \{1,2,\cdots,M\}                  \Bigg\}.
\end{align}
The transmitted signal further undergoes a slowly varying flat Rayleigh fading channel ${g}\!\!\sim\!\!{\mathcal{CN}}(0,\lambda,0)$. Moreover, the receiver further induces an additive distortion ${\eta_{\mathrm{r}}}$, resulting from the  non-linear transfer function of low noise amplifier, band-pass filters, image rejection low pass filter, analog-to-digital converter. It is important to highlight that the receiver distortions are in addition to the conventional thermal noise at the receiver.
\begin{equation} \label{eq1}
y = \sqrt {\alpha}\, g \,\left( {x_m + \eta_{\mathrm{t}}} \right) + \eta _{\rm r}  + w; \quad m \in \{ 1,2, \cdots ,M\},
\end{equation}
where $\alpha$ is the transmitted power. The \ac{AWGN} $w$ and receiver  \ac{HWD} $\eta_{\mathrm{r}}$
are distributed as $w \sim \mathcal{CN}(0, \sigma_w^2 , 0 )$ and $\eta _{\rm r} \sim \mathcal{CN}(0, \alpha |g|^2 \kappa_{\rm r} , \alpha g^2 \tilde{\kappa_{\rm r}})$. 
The additive Gaussian distortion model for the aggregate residual \ac{RF} distortions is backed by various theoretical investigations and measurement results (see, e.g.,\cite{wenk2010mimo,zetterberg2011experimental,
bjornson2013capacity,boulogeorgos2016energy,
bjornson2013new,xia2015hardware,suzuki2008transmitter,
studer2010mimo,duy2015proactive,bjornson2014massive} and references therein). This can also be motivated analytically by the central limit theorem. Furthermore, the improper nature of these distortions is motivated by the imbalance between in-phase and quadrature-phase branches in the up-conversion and down-conversion phases \cite{javed2016impact}.
\begin{lemma}[Aggregate effect of transceiver distortions \cite{javed2017asymmetric,schenk2008rf}]
For  the following generalized received signal model 
\begin{equation} \label{eq2}
y = \sqrt \alpha  g{x_m} + z ;\quad m \in \{ 1,2, \ldots ,M\},
\end{equation}
where $z \triangleq \sqrt \alpha  g\eta  + w$ is the r.v. representing the aggregate effect of transceiver distortions, $\eta  \sim \mathcal{CN}(0, \kappa, \tilde{\kappa})$, ${\kappa }= {\kappa _{ \mathrm {t} }  + \kappa _{\rm r} } $ and $\tilde{\kappa} = \tilde{\kappa}_{\rm{t}} + \tilde{\kappa}_{\rm{r}}  $,  the aggregate interference can be modeled as improper noise, i.e.,  $z \sim \mathcal{CN}\left( {0,\alpha {{\left| g \right|}^2}\kappa  + \sigma _w^2,\alpha {g^2}\tilde \kappa } \right) $. Moreover, the variance of $z_{\rm I}$ and $z_{\rm Q}$ are given in \eqref{eq3} and \eqref{eq4}, respectively, as
\begin{equation}\label{eq3}
\sigma_{\rm I}^2 =  \frac{{\alpha {{\left| g \right|}^2}\kappa  + \sigma _w^2 + \alpha {\Re}\left( {{g^2}\tilde \kappa } \right)}}{2},
\end{equation}
\begin{equation}\label{eq4}
\sigma_{\rm Q}^2 = \frac{{\alpha {{\left| g \right|}^2}\kappa  + \sigma _w^2 - \alpha {\Re}\left( {{g^2}\tilde \kappa } \right)}}{2}.
\end{equation}
Furthermore, the non-zero pseudo-variance $\tilde \sigma_z^2$ motivates us to evaluate the correlation between $z_{\rm I}$ and $z_{\rm Q}$ using the correlation coefficient $\rho_z$ as
\begin{equation}\label{eq5}
\rho_z = \frac{{\alpha {\Im}\left( {{g^2}\tilde \kappa } \right)}}{{\sqrt {{{\left( {\alpha {{\left| g \right|}^2}\kappa  + \sigma _w^2} \right)}^2} - {{\left( {\alpha {\Re}\left( {{g^2}\tilde \kappa } \right)} \right)}^2}} }}.
\end{equation}
Proof of \eqref{eq3}-\eqref{eq5} is presented in Appendix \ref{AppendA}. 
\end{lemma}
It is important to note that \eqref{eq2} reduces to the conventional signal model $y = \sqrt \alpha  g{x_m}  + w$  in case of ideal hardware, i.e., $\kappa = 0$, which is induced by imposing $\kappa_{\rm t} = \kappa_{\rm r}=0 \:$and also $\tilde{\kappa} = 0$, which is deduced from Definition 2. 

 \ac{HWD} can leave drastic effects on the system performance as they raise the noise floor. Although, the entropy loss of improper noise is less than the proper noise but it is difficult to tackle. It requires some meticulously designed improper signaling like \ac{IGS} for effective mitigation. However, \ac{IGS} is difficult to implement because of the unbounded peak-to-average power ratio and high detection complexity \cite{santamaria2018information,javed2020journey}. Therefore, researchers resort to the finite discrete \ac{AS}  schemes obtained by GS.

We propose \ac{PS} as another way to realize AS in order to effectively dampen the deteriorating effects of improper  \ac{HWD}. \ac{PS} aims to design non-uniform symbol probabilities for a higher order QAM to minimize \ac{BER}  offering more degrees of freedom and adaptive rates. In the following section, we carry out the error probability analysis of the adopted system which lays foundation for the proposed \ac{PS} design. 

\subsection{Optimal Receiver}
Conventional systems with Gaussian interference employ least-complex receivers with either minimum Euclidean or maximum likelihood detectors. However, such receivers cannot accommodate the unequal symbol probabilities and improper noise. Therefore, the optimal detection in the presented scenario can only be achieved by the \ac{MAP} detector at the expense of increased receiver complexity. Considering the improper Gaussian \ac{HWD} and the non-uniform priors of the constellation symbols, the optimal \ac{MAP}  detection is given by
\begin{equation}\label{eq6}
\hat m_{\rm PS} = \mathop {\arg \max }\limits_{1 \le m \le M}  \quad p_X(x_m) {f_{{Y_{\rm I}},{Y_{\rm Q}}|X,g}}\left( {{y_{\rm I}},{y_{\rm Q}}|{x_m},g} \right),
\end{equation} 
\begin{floatEq}
\begin{equation}\label{eq7}
{f_{{Y_{\rm I}},{Y_{\rm Q}}|X,g}}\left( {{y_{\rm I}},{y_{\rm Q}}|x_m,g} \right) = \frac{1}{{2\pi {\sigma _{\rm I}}{\sigma_{\rm Q}}\sqrt {1 - {\rho_z^2}} }}\exp \left\{ { - \frac{1}{{2\left( {1 - {\rho_z ^2}} \right)}}\left[ \begin{array}{l}
\frac{{{{\left( {{y_{\rm I}} - \sqrt \alpha  {\Re}\left( {g{x_m}} \right)} \right)}^2}}}{{\sigma _{\rm I}^2}} + \frac{{{{\left( {{y_{\rm Q}} - \sqrt \alpha  {\Im}\left( {g{x_m}} \right)} \right)}^2}}}{{\sigma _{\rm Q}^2}} +  \\
-\frac{{2\rho_z \left( {{y_{\rm I}} - \sqrt \alpha  {\Re}\left( {g{x_m}} \right)} \right)\left( {{y_{\rm Q}} - \sqrt \alpha  {\Im}\left( {g{x_m}} \right)} \right)}}{{{\sigma _{\rm I}}{\sigma _{\rm Q}}}}
\end{array} \right]} \right\}.
\end{equation}
\end{floatEq}
where  ${f_{{Y_{\rm I}},{Y_{\rm Q}}|X,g}}\left( {{y_{\rm I}},{y_{\rm Q}}|{x_m},g} \right)$ is the conditional Gaussian \ac{PDF} of $y$ representing \ac{ML} function given $x_m$ and $g$, as expressed in \eqref{eq7} at the top of next page.
\section{Error Probability Analysis} \label{SecIII}
Considering the non-uniform priors and improper noise, the error probability analysis is carried out based on the optimal \ac{MAP}  detector presented in Section \ref{SecII}. Symbol error probability $P_{\rm s}$ is the accumulated error probability of all symbols with respect to their prior probabilities and is given as
\begin{equation}\label{eq8}
{P_{\rm s}} = \sum\limits_{m = 1}^{M}  { {p_m} \Pr \left( e|x_m \right)},
\end{equation}
where $\Pr \left( e|x_m \right)$ is the probability of an error event given symbol $x_m$ was transmitted. 
In order to yield a tractable and simplified analysis especially for higher order modulation schemes,  $P_{\rm s}$ can be upper bounded as
\begin{equation} \label{eq10}
{P_{\rm s}} \le \sum\limits_{m = 1}^{M} {\sum\limits_{ n=1 \atop n \ne m}^{M} {{p _m} P_{mn}  }},
\end{equation}
where, $P_{mn}$ 
is the pairwise error probability (PEP), which represents the probability of deciding ${x_n}$ given ${x_m}$ was transmitted, ignoring all the other symbols in the constellation \cite{john2008digital}.
The PEP can be evaluated using the \ac{MAP}  rule in \eqref{eq6} as
\begin{multline}\label{eq11}
    {P}_{mn} = \Pr \left\{ {p_m}\,{f_{{Y_{\rm{I}}},{Y_{\rm{Q}}}|X,g}}\left( {{y_{\rm{I}}},{y_{\rm{Q}}}|{x_m},g} \right) \le  {p_n}  \right. \\
    \left. \times {f_{{Y_{\rm{I}}},{Y_{\rm{Q}}}|X,g}}\left( {{y_{\rm{I}}},{y_{\rm{Q}}}|{x_n},g} \right) 
    \right\}.
\end{multline}
By substituting the conditional probability from  \eqref{eq7} in \eqref{eq11} and after some mathematical simplifications, the PEP can be written as in  \eqref{eq12}, shown in the next page. Now, we find the in-phase and quadrature-phase components of the received signal $y$ for a given transmitted symbol $x_m$ as follows
\begin{floatEq}
\begin{equation}\label{eq12}
{P}_{mn} = \Pr \left\{ 2\left( {1 - \rho _z^2} \right)\ln \left( {\frac{{{p_m}}}{{{p_n}}}} \right) \le \left[ \begin{array}{l}
\frac{{{{\left( {{y_{\rm{I}}} - \sqrt \alpha  \Re \left( {g{x_m}} \right)} \right)}^2} - {{\left( {{y_{\rm{I}}} - \sqrt \alpha  \Re \left( {g{x_n}} \right)} \right)}^2}}}{{\sigma _{\rm{I}}^2}} + \frac{{{{\left( {{y_{\rm{Q}}} - \sqrt \alpha  \Im \left( {g{x_m}} \right)} \right)}^2} - {{\left( {{y_{\rm{Q}}} - \sqrt \alpha  \Im \left( {g{x_n}} \right)} \right)}^2}}}{{\sigma _{\rm{Q}}^2}} +  \\
+\frac{{2{\rho _z}\left( {{y_{\rm{I}}} - \sqrt \alpha  \Re \left( {g{x_n}} \right)} \right)\left( {{y_{\rm{Q}}} - \sqrt \alpha  \Im \left( {g{x_n}} \right)} \right) - 2{\rho _z}\left( {{y_{\rm{I}}} - \sqrt \alpha  \Re \left( {g{x_m}} \right)} \right)\left( {{y_{\rm{Q}}} - \sqrt \alpha  \Im \left( {g{x_m}} \right)} \right)}}{{{\sigma _{\rm{I}}}{\sigma _{\rm{Q}}}}}
\end{array} \right] \right\}.
\end{equation}
\end{floatEq}
\begin{equation}\label{eq13}
{y_{\rm I}} = \sqrt \alpha  \Re \left( {g{x_m}} \right) + {z_I},
\end{equation}
and
\begin{equation}\label{eq14}
{y_{\rm Q}} = \sqrt \alpha  \Im \left( {g{x_m}} \right) + {z_Q},
\end{equation}
respectively.
Then, we substitute ${y_{\rm I}}$ and ${y_{\rm Q}}$ in \eqref{eq12}, which can be further simplified obtaining,
\begin{equation}\label{eq15}
{P}_{mn} = \Pr \left\{ 
\psi  \ge 2\left( {1 - \rho _z^2} \right)\ln \left( {\frac{{{p_m}}}{{{p_n}}}} \right)  + \alpha {\gamma _{mn}}  \right\},
\end{equation}
where 
\begin{equation}\label{eq17}
{\gamma _{mn}} \triangleq \frac{{{\xi_{mn}}_I^2}}{{\sigma _{\rm{I}}^2}} + \frac{{{\xi_{mn}}_Q^2}}{{\sigma _{\rm{Q}}^2}} - \frac{{2{\rho _z}{{\xi_{mn}}_I}{{\xi_{mn}} _Q}}}{{{\sigma _{\rm{I}}}{\sigma _{\rm{Q}}}}},
\end{equation}
with ${\xi_{mn}} = g\, {d_{mn}} = g\left( {{x_m} - {x_n}} \right)$ representing the distance between $m^{\rm th}$ and $n^{\rm th}$  symbol with channel coefficient $g$, and $\psi$ is obtained by the superposition of ${z_I}$ and ${z_Q}$ as
\begin{equation}\label{eq16}
\psi  \!\!= \!\!2\sqrt \alpha  {\rho _z}\left[ {\left( {\frac{{{{\xi_{mn}}_Q}}}{{{\sigma _{\rm{I}}}{\sigma _{\rm{Q}}}}}\!\! -\!\! \frac{{{{\xi_{mn}}_I}}}{{{\rho _z}\sigma _{\rm{I}}^2}}} \right){z_I} \!\!+ \!\!\left( {\frac{{{{\xi_{mn}}_I}}}{{{\sigma _{\rm{I}}}{\sigma _{\rm{Q}}}}}\!\! - \!\!\frac{{{{\xi_{mn}}_Q}}}{{{\rho _z}\sigma _{\rm{Q}}^2}}} \right){z_Q}} \right].
\end{equation}
Clearly, $\psi$ is another zero mean Gaussian random variable with variance $\sigma _\psi ^2$ expressed as
\begin{equation}
\sigma _\psi ^2 = 4 \left( {1 - \rho _z^2} \right)\alpha{\gamma _{mn}}.
\end{equation}
Conclusively, ${P}_{mn}$ is the complementary cumulative distribution function of $\psi$ and is given as
\begin{equation}\label{eq19}
{P}_{mn} = \mathcal{Q} \left( {\frac{{2\left( {1 - \rho _z^2} \right)\ln \left( {\frac{{{p_m}}}{{{p_n}}}} \right) + \alpha {\gamma _{mn}}}}{2{\sqrt {\left( {1 - \rho _z^2} \right)\alpha {\gamma _{mn}}} }}} \right).
\end{equation}
Substituting the PEP derived in \eqref{eq19} to \eqref{eq10} along with the gray mapping assumption yields the following bound on \ac{BER} 
 \begin{equation}\label{eq20}
{P_{\rm b}} \! \le \!{{\rm{P}}_{\rm b}^{\rm UB}}\!\triangleq\!  \frac{1}{{{{\log }_2} \left( M \right)}}\!\!\sum\limits_{m = 1}^M {\sum\limits_{{n = 1}\atop
{n \ne m}}^M\! {{p_m}  \mathcal{Q}\! \left( {{\beta _{mn}}}\! \ln \left( {\frac{{{p_m}}}{{{p_n}}}} \right)\!+
\!\frac{1}{2{{\beta _{mn}}} } \right)} },
\end{equation}
where ${\beta _{mn}} \triangleq \sqrt{1 - \rho _z^2}/ \sqrt {\alpha {\gamma _{mn}}}$. The \ac{BER}  expression depends on the size of the constellation,
prior probabilities of all the symbols, power budget,
mutual distances between the transmitted and received erroneous symbols under Rayleigh fading, and  \ac{HWD} statistical characteristics. 

In contrast to the monotonically decreasing \ac{BER}  for the ideal systems, the \ac{BER}  saturates after a specific \ac{SNR}  in the hardware-distorted transceivers.
In this regard, we carry out the asymptotic analysis of the bit error probability to quantify the error floor as high \ac{SNR}. Let us set 
\begin{equation} \label{eqUp}
  \Upsilon \triangleq  {1 - \frac{{{{\left( {{\Im}\left( {{g^2}\tilde \kappa } \right)} \right)}^2}}}{{\left( {{{\left| g \right|}^4}{\kappa ^2}} \right) - {{\left( {{\Re}\left( {{g^2}\tilde \kappa } \right)} \right)}^2}}}},
\end{equation}
the error floor can be upper bounded from  \eqref{eq20} as in \eqref{eqAsymp}. We can see that the error floor depends on the adopted $M$-ary constellation, channel coefficient,  \ac{HWD} statistical characteristics, and symbol probabilities.

\begin{floatEq}
\begin{equation} \label{eqAsymp}
\mathop {\lim }\limits_{\alpha  \to \infty } 
{P_{\rm b}}  \le   \frac{1}{{{{\log }_2} \left( M \right)}}\!\!\sum\limits_{m = 1}^M {\sum\limits_{{n = 1}\atop
{n \ne m}}^M\! {{p_m} Q\left( {\frac{{2 \Upsilon \ln \left( {\frac{{p_m}}{{p_n}}} \right) + \left( {\frac{{2{\Re}{{\left( {g{d_{mn}}} \right)}^2}}}{{{{\left| g \right|}^2}\kappa  + {\Re}\left( {{g^2}\tilde \kappa } \right)}} + \frac{{2{\Im}{{\left( {g{d_{mn}}} \right)}^2}}}{{{{\left| g \right|}^2}\kappa  - {\Re}\left( {{g^2}\tilde \kappa } \right)}} - 2\frac{{{\Re}\left( {g{d_{mn}}} \right){\Im}\left( {g{d_{mn}}} \right){\Im}\left( {{g^2}\tilde \kappa } \right)}}{{{{\left( {{{\left| g \right|}^2}\kappa } \right)}^2} - {{\left( {{\Re}\left( {{g^2}\tilde \kappa } \right)} \right)}^2}}}} \right)}}{{\sqrt {4\Upsilon \left( {\frac{{2{\Re}{{\left( {g{d_{mn}}} \right)}^2}}}{{{{\left| g \right|}^2}\kappa  + {\Re}\left( {{g^2}\tilde \kappa } \right)}} + \frac{{2{\Im}{{\left( {g{d_{mn}}} \right)}^2}}}{{{{\left| g \right|}^2}\kappa  - {\Re}\left( {{g^2}\tilde \kappa } \right)}} - 2\frac{{{\Re}\left( {g{d_{mn}}} \right){\Im}\left( {g{d_{mn}}} \right){\Im}\left( {{g^2}\tilde \kappa } \right)}}{{{{\left( {{{\left| g \right|}^2}\kappa } \right)}^2} - {{\left( {{\Re}\left( {{g^2}\tilde \kappa } \right)} \right)}^2}}}} \right)} }}} \right)} }.
\end{equation} 
\end{floatEq}

\section{Proposed Probabilistic Signaling Design}\label{ssec:PS}

We aim to design the non-uniform symbol probabilities, which minimize the \ac{BER}  of the adopted system suffering from  \ac{HWD}. The optimization is carried out given power and rate constraints. The rate of the conventional QAM with uniform symbol probabilities and modulation order $M_{\rm u}$ is fixed, i.e., $R=\log_2(M_{\rm u})$. However, we seek the maximum benefits of \ac{PS} by allowing a higher-order modulation with $M_{\rm nu} > M_{\rm u}$, where $M_{\rm nu}$ is the modulation order of the constellation with non-uniform probabilities $\bf p$.
Thus, the rate of this scheme can be designed such that $R \triangleq {\rm H}({\bf p})\geq \log_2(M_{\rm u})$, rendering more design flexibility and hence is capable of reducing the \ac{BER}. \ac{PS} is capable of changing the transmission rate by changing the symbol distribution for a fixed modulation order, unlike uniform signaling, which needs to change the modulation scheme's order to change the rate for uncoded communications.

After designing the symbol probabilities, we can implement \ac{PS} by using distribution matching at the transmitter to map uniformly distributed input bits to $M_{\rm nu}$-QAM/PSK symbols \cite{SchBoc:16,schulte2014zero,dia2018compressed}.
Moreover, they can be detected using the proposed \ac{MAP}  detector \eqref{eq6} 
at the receiver that incorporates the prior symbol distribution. In the following, we formulate the \ac{PS} design problem and propose an algorithm to obtain the non-uniform symbol probabilities followed by some toy examples.
\subsection{Problem Formulation}
The probability vector
 ${\mathbf{p}} \triangleq [p_{1},{p_2}, \ldots ,{p_{M_{\rm nu}}}]$, containing probabilities
 of the symmetric $M_{\rm nu}-$QAM/PSK modulated symbols with $M_{\rm nu}> M_{\rm u}$ \footnote{For $M_{\rm nu}= M_{\rm u}$, the distribution should be uniform to satisfy the rate constraint because uniform signaling has the largest entropy.}, is designed to minimize the upper bound on the \ac{BER}  derived in \eqref{eq20}. In particular, we formulate the problem as
 
\begin{subequations}\label{eq.P1}
\begin{alignat}{2}
\textbf{P1}:\quad &\!\!\!\!\underset{{\bf p}\in \mathbb{S}}{\text{minimize }}
&&{{\rm{P}}_{\rm b}^{\rm UB}}\left( {\mathbf{p}}  \right) \\
&\!\!\!\! \text{subject to} \quad
& & \sum\limits_{m = 1}^{M_{\rm nu}} {{{\left| {{x_m}} \right|}^2}{p_m} \le 1}, \label{eq.pc} \\
&&& {\rm H}({\mathbf{p}})  \geq \log_2\left({M_{\rm u}}\right), \label{eq.rc}
\end{alignat}
\end{subequations}
where \eqref{eq.pc} and \eqref{eq.rc} represent the average power and rate constraints, respectively, and ${\rm H}({\mathbf{p}})$ is the source entropy, which  represents  the transmitted rate in terms of bits per symbol per channel use and is defined as
\begin{equation}\label{eq21}
{\rm H}({\mathbf{p}}) \triangleq  \sum\limits_{m = 1}^{{M_{\rm nu}}} -{p_m}\, \log_2\left({p_m} \right).
\end{equation}

The concave nature of information entropy in \eqref{eq.rc} renders a convex constraint in $\mathbf{p}$  and the rate fairness is justified based on the trade off between \ac{BER}  minimization and rate maximization, while satisfying a minimum rate. Therefore, the idea is to employ a higher order non-uniformly distributed $M_{\rm nu}-$QAM/PSK as compared to a lower order uniformly distributed $M_{\rm u}-$QAM/PSK with same energy and at least the same rate to minimize \ac{BER}.

\subsection{Optimization Framework}
The optimization problem \textbf{P1} \eqref{eq.P1} is a non-convex optimization problem owing to the non-convex objective function even though all the constraints are convex. Therefore, we propose successive convex approximation approach to tackle it. We begin by approximating 
$ {{\rm{P}}_{\rm b}^{\rm UB}}\left( {\mathbf{p}}  \right) $ with its first order Taylor series approximation.

First order Taylor series approximation of a function $f \left( x \right)$ around a point $ x^{(k)}$ is given as
\begin{equation}\label{eq22}
\tilde f\left( {x,x^{(k)} } \right) \approx f\left( {{{x}^{\left( k \right)}}} \right) + \nabla_x f\left( {{{x}^{\left( k \right)}}} \right)\left( {{x} - {{x}^{\left( k \right)}}} \right).
\end{equation}
Thus, we need to compute $\nabla_{{\mathbf{p}}} {\rm{P}}_{\rm b}^{\rm UB}$ and evaluate it at ${{{\mathbf{p}}}}^{\left( k \right)}$ to compute ${\tilde{\rm{P}}_{\rm b}^{\rm UB}}\left( {\mathbf{p}},{\mathbf{p}}^{(k)}   \right)$.
\begin{equation}\label{eq23}
 \nabla_{{\mathbf{p}}} {\rm{P}}_{\rm b}^{\rm UB} = \left[  \frac{{\partial {\rm{P}}_{\rm b}^{\rm UB}}}{{\partial {p_1}}} \quad  \frac{{\partial {\rm{P}}_{\rm b}^{\rm UB}}}{{\partial {p_2}}} \quad \ldots \quad  \frac{{\partial {\rm{P}}_{\rm b}^{\rm UB}}}{{\partial {p_{M_{\rm nu}}}}}  \right].
\end{equation}
In order to compute  ${{\partial {\rm{P}}_{\rm b}^{\rm UB}}}/{{\partial {p_t}}}$, we rewrite \eqref{eq20} as
\begin{equation}\label{eq24}
{{\rm{P}}_{\rm b}^{\rm UB}} = \frac{1}{{{{\log }_2}\left( M_{\rm nu} \right)}}\sum\limits_{m = 1}^{M_{\rm nu}}  \sum\limits_{{n = 1} \atop
{n \ne m}}^{M_{\rm nu}}  {p_m}{\int\limits_{\Omega_{mn}}^\infty  {\frac{{{e^{ - \frac{{{u^2}}}{2}}}}}{{\sqrt {2\pi } }}du} },
\end{equation}
where 
\begin{equation}
    \Omega_{mn} = {{\beta _{mn}}\ln \left( {\frac{{{p_m}}}{{{p_n}}}} \right) + \frac{1}{{2{\beta _{mn}}}}}.
\end{equation}
From \eqref{eq24} and by applying the Leibniz integral rule, we get
\begin{align}\label{eq26}
\frac{{\partial {\rm P_b^{\rm UB}}}}{{\partial {p_t}}} \le &  \frac{1}{\log_2\left(M_{\rm nu}\right)} \sum\limits_{{n = 1,} \atop {{n \ne t,} \atop {m = t}}}^{M_{\rm nu} } {}  \left({\cal Q}\left( \Omega_{mn}  \right) - \frac{{{\beta _{mn}}}}{{\sqrt {2\pi } }}{e^{ - \frac{\Omega_{mn}^2 }{2}}}\right)  \nonumber \\
& +\frac{1}{\log_2\left(M_{\rm nu}\right)} \sum\limits_{{m = 1,}\atop {{m \ne t,} \atop {n = t}}}^{M_{\rm nu} } {} {\frac{{{\beta _{mn}}{p_m}}}{{\sqrt {2\pi } {p_n}}}{e^{ - \frac{\Omega_{mn}^2 }{2}}}}. 
\end{align}
Now, ${{\rm{P}}_{\rm b}^{\rm UB}}$ can be approximated from   \eqref{eq22}, \eqref{eq23},  and \eqref{eq26} using first order Taylor series expansion around an initial probability vector ${\mathbf{p}}^{(k)}$ as
\begin{equation}\label{eq27}
 {\tilde{\rm{P}}_{\rm b}^{\rm UB}} \! \left( {\mathbf{p}},{\mathbf{p}}^{(k)}   \! \right) \! \triangleq\!  {{\rm{P}}_{\rm b}^{\rm UB}}\! \left( {\mathbf{p}}^{(k)}   \! \right)\!+\! \nabla_{{\mathbf{p}}}{{\rm{P}}_{\rm b}^{\rm UB}} \! \left( \!{\mathbf{p}}^{(k)}  \! \right) \!\left( \!{\mathbf{p}} \!-\! {\mathbf{p}}^{(k)}  \! \right).
\end{equation}
\begin{algorithm}[!t]
\caption{Successive Convex Programming}\label{Algo:SCP}
\begin{algorithmic}[1]
\State \textbf{Initialize}  $i \gets 0$, $\epsilon \gets \infty$   and   \textbf {Set}   tolerance $\delta$
\State \textbf{Choose} feasible starting point $ {\mathbf{p}}^{(i)}$
\While {$\epsilon \ge \delta$}
\State Evaluate $ {\tilde{\rm{P}}_{\rm b}^{\rm UB}}\left( {\mathbf{p}},{\mathbf{p}}^{(i)}   \right)$
\State  Solve \textbf{P1a} and obtain  ${\mathbf{p}}$ using ${\mathbf{p}}^{(i)}$
\State   ${{\mathbf{p}}}^{(i+1)} \gets {\mathbf{p}}$
\State Update ${\epsilon \gets \left\| {\mathbf{p}}^{(i+1)}- {\mathbf{p}}^{(i)}\right\|}$
\State   ${i \gets i+1}$
\EndWhile
\State   ${\mathbf{p}}^* \gets {\mathbf{p}}^{i+1}$
\State  $ \rm{P}_{\rm b}^* \leq \rm{P}_{\rm b}^{\rm UB}\left( \mathcal{P^*}\right) $
\end{algorithmic}
\end{algorithm}
Successive convex programming minimizes \textbf{P1} by iteratively solving its convex approximation \textbf{P1a} as presented in Algorithm \ref{Algo:SCP}.
\begin{subequations}\label{eq.P1a}
\begin{alignat}{2}
\textbf{P1a}:\quad &\!\!\!\!\underset{{\bf p}\in \mathbb{S}}{\text{minimize }}
&&{\tilde{\rm{P}}_{\rm b}^{\rm UB}}\left( {\mathbf{p}},{\mathbf{p}}^{(k)}   \right) \\
&\!\!\!\! \text{subject to} \quad
& & \sum\limits_{m = 1}^{M_{\rm nu}} {{{\left| {{x_m}} \right|}^2}{p_m} \le 1},  \\
&&& {\rm H}({\mathbf{p}})  \geq \log_2\left({M_{\rm u}}\right), 
\end{alignat}
\end{subequations}
It begins with the initiation of counter $i$, stopping criteria $\epsilon$ and the stopping threshold $\delta$. Secondly, we choose some feasible PMF set ${\mathbf{p}}^{(i)} \in \mathbb{S}$ which satisfies the constraints  \eqref{eq.pc} and \eqref{eq.rc}.  The while loop starts by evaluating the approximation ${\tilde{\rm{P}}_{\rm b}^{\rm UB}}\left( {\mathbf{p}},{\mathbf{p}}^{(i)}   \right)$ around ${\mathbf{p}}^{(i)}$. 

The convex problem \textbf{P1a} is solved using the 
Karush Kuhn Tucker (KKT) conditions derived in Appendix \ref{AppendC} to obtain the optimal probabilities for \textbf{P1a}\cite{boyd2004convex}. The solution obtained in  
this iteration is updated as ${\mathbf{p}}^{(i+1)} $ and is used to evaluate the stopping criteria ${\epsilon \gets \left\| {\mathbf{p}}^{(i+1)}- {\mathbf{p}}^{(i)}\right\|}$ as shown in Algorithm \ref{Algo:SCP}. The loop ends when the change in two subsequent solution parameters in terms of the $\ell_{2}$ norm is less than a predefined threshold $\delta$. Once the stopping criteria is attained, the solution parameters ${\mathbf{p}}^{(*)}$ are guaranteed to render a \ac{BER} $\rm{P}_{\rm b}^* $ which will be lower than the bound  $\rm{P}_{\rm b}^{\rm UB}\left( \mathcal{P^*}\right) $.

\begin{figure}[t]
    \centering
   \includegraphics[width=3.5in]{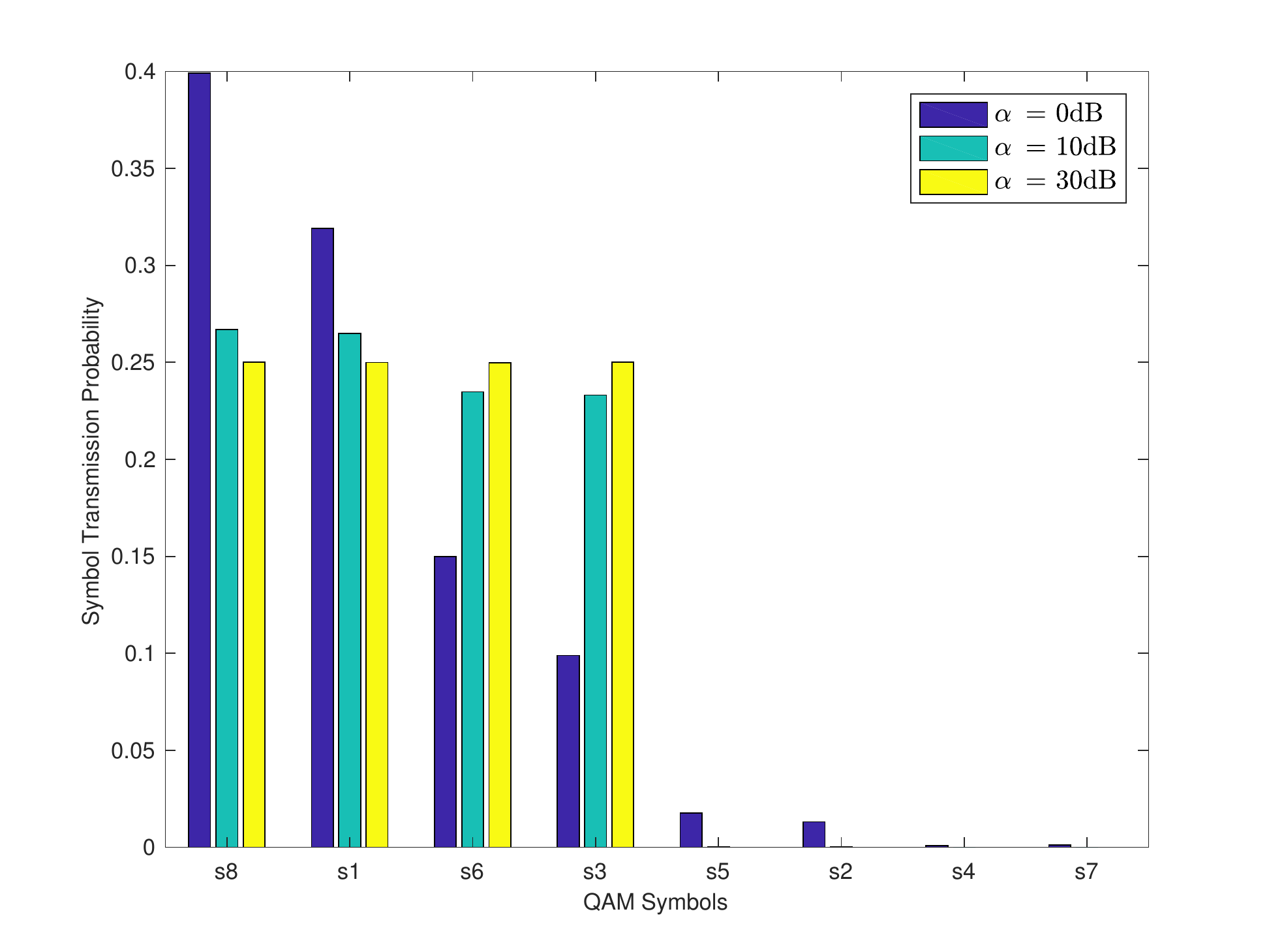}
    \caption{ $8$-QAM probability distribution for three \ac{SNR}  levels at $\eta=0.99$ and Rate = $2$ bits/symbol. }
    \label{fig:PS_3SNR}
\end{figure}
\subsection{Toy Examples}
\begin{figure}[t]
    \centering
   \includegraphics[width=3.5in]{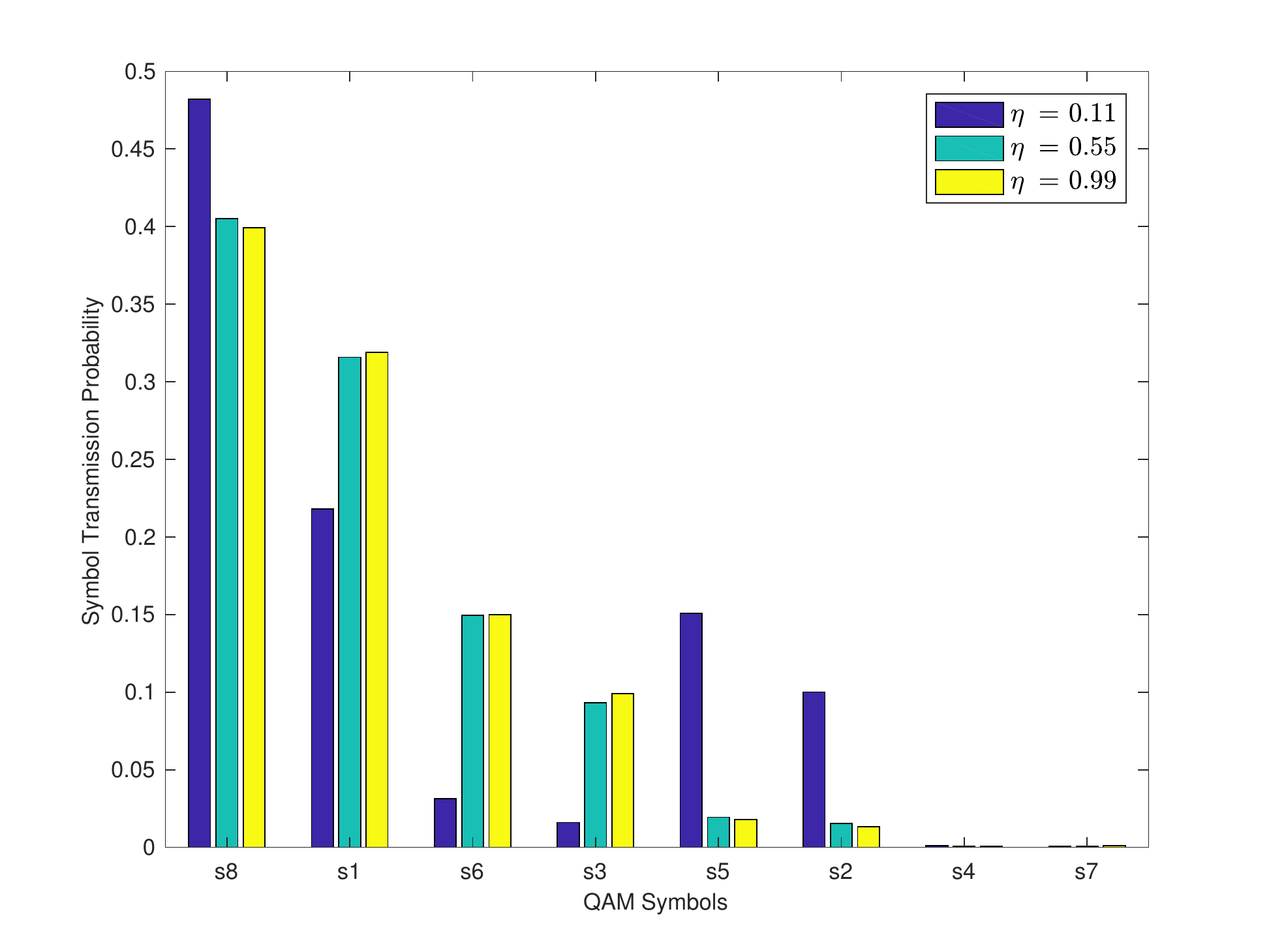}
    \caption{ $8$-QAM probability distribution for three  \ac{HWD} levels at $\alpha=0$~dB and Rate = $2$~bits/symbol. }
    \label{fig:PS_3HWD}
\end{figure}

A comprehensive illustration of probabilistically shaped $M_{\rm nu}=8$-QAM with a $2$ bits/symbol rate constraint, corresponding to $M_{\rm u}=4$, is presented in Fig. \ref{fig:PS_3SNR} and Fig. \ref{fig:PS_3HWD}. The relation between prior probabilities and different \ac{SNR}  values is presented in Fig.  \ref{fig:PS_3SNR}. Clearly, the probability distribution is quite random for lower \ac{SNR}  level such as $\alpha=0$~dB. However, it starts adopting uniform distribution of $0.25$ for four of it's symbols, i.e., s1, s3, s6, and s8 while zero probabilities for the rest four symbols. This technique provides lower \ac{BER}  while maintaining $2$~bits/symbol rate for a fair comparison with traditional $4$-QAM. 
Interestingly, it achieves a lower \ac{BER}  by transmitting half of the symbols which are not the nearest neighbors. It is important to highlight that the proposed approach achieves this performance with the same power budget and transmission rate. 

Another example illustrates the trend of probabilitic shaping for $8$-QAM constellation at lower \ac{SNR}  level (keeping in mind that it assigns the uniform probabilities to four symbols at high \ac{SNR}  levels). The trend for lower  \ac{HWD} level such as $\eta=0.11$ is quite random. However, it follows a decreasing probability trend for middle to higher  \ac{HWD} levels. Intuitively, it assigns higher probabilities to the symbols with least power and lower probabilities to the symbols with higher powers. 
This trend decreases the \ac{BER}  while maintaining the average power constraint.

\section{Hybrid Shaping where Conventional meets State-of -the-Art }
In this section, we increase the \acs{AS} design flexibility by allowing joint \acs{GS} and \acs{PS}, which we call it here \acs{HS}, to improve the underlying communication system performance further. Throughout the design procedure, \acs{HS} transforms the equally spaced uniformly distributed QAM/PSK symbols to unequally spaced symbols in a geometric envelope with non-uniform prior distribution. Thus, \acs{HS}   aims to optimize the symbol probabilities (i.e., \acs{PS}) and some spatial shaping parameters for the constellation (i.e., \acs{GS}). 
\subsection{Hybrid Shaping Parameterization}
Apart from the non-uniform priors, consider the asymmetric transmit symbol ${\bf v}_m = \left[ {v}_{mI} \quad  {v}_{mQ} \right]^{\rm T} $ resulting from the \ac{GS} on the conventional baseband symmetric $M$-QAM/$M$-PSK symbol ${\bf x}_m = \left[ {x}_{mI} \quad {x}_{mQ} \right]^{\rm T} $ as $\V{v}_m = \M{AR}\V{x}_m$, where
\begin{equation}\label{eq28}
{\bf{A}} \left(\zeta \right)= \left[ {\begin{array}{*{20}{c}}
{ \sqrt{1+\zeta} }&{0}\\
{0}&{ \sqrt{1-\zeta}}
\end{array}} \right],
\end{equation}
with translation parameter $\zeta \in \left( 0,1\right)$. Furthermore, the rotation is given by
\begin{equation}\label{eq29}
{\bf{R}}\left(\theta \right)= \left[ {\begin{array}{*{20}{c}}
{\cos\left( \theta \right)}&{-\sin\left( \theta \right)}\\
{\sin\left( \theta \right)}&\quad{ \cos\left( \theta \right)}
\end{array}} \right],
\end{equation}
with rotation angle $\theta \in \left(0,\mu\, \pi/2 \right)$ for some constant $\mu$. Uniformly distributed symmetric $M$-QAM constellation has a rotation symmetry of $n\,\pi/2, n\in \mathcal{Z}^{+}$ rendering $\mu =n$ to be good choice for GS. However, non-uniformly distributed $M$-QAM  constellation can only be rotationally symmetric after $2n\pi$, thus $\mu = 4n$ is suitable for HS. This technique renders non-uniformly spaced symbols in a parallelogram envelop. It is important to highlight that this transformation preserves the power requirement. Power invariance of the rotation is a well known fact in the literature  \cite{john2008digital}. However, the wisdom behind the structure of $\M{A} \left(\zeta \right)$ is unfolded in the following theorem. 
\begin{remark}\label{remark:powertranslation}
GS parameterization using translation matrix ${\M{A}} \left(\zeta \right) $ preserves the power invariance of a complex random variable and inculcates asymmetry/improperness with the circularity coefficient $\zeta$.
\end{remark}
\begin{proof}
The proof  is presented in Appendix \ref{AppendB}. Furthermore, the generalization of the same concept to the symmetric discrete constellations such as $M$-QAM and $M$-PSK is also described in Appendix \ref{AppendB}.
\end{proof}
\subsection{Optimal Receiver}
 The optimal receiver for hybrid shaped \ac{AS}  is also a \ac{MAP}  detector as derived in \eqref{eq6}, but with a modified reference constellation $v_m$ in place of $x_m$ for all  $m \in \{1,2,\cdots, M_{\rm nu}\}$. More precisely, the detected symbol, $\hat m_{\rm HS}$, is the one that maximizes the posterior distribution, i.e.,
\begin{equation}\label{eq30}
\hat m_{\rm HS} = \mathop {\arg \max }\limits_{1 \le m \le M_{\rm nu}}  \quad p_V(v_m) {f_{{Y_{\rm I}},{Y_{\rm Q}}|V,g}}\left( {{y_{\rm I}},{y_{\rm Q}}|{v_m},g} \right),
\end{equation}  
 where, ${f_{{Y_{\rm I}},{Y_{\rm Q}}|V,g}}\left( {{y_{\rm I}},{y_{\rm Q}}|{v_m},g} \right)$ is similar to \eqref{eq7} by replacing all appearances if $x_m$ with $v_m$ for all  $m \in \{1,2,\cdots, M_{\rm nu}\}$. It is worth noting that non-uniform prior probabilities are inculcated in the detection process using \ac{MAP}  detector in place of \ac{ML} detector. Moreover, the geometrically shaped symbols are taken from a modified symbol constellation. Hence, this requires updating the reference constellation for appropriate detection. 
\subsection{Error Probability} 
 \ac{HS}  follows the same \ac{BER}  bound as derived in \eqref{eq20} but with modified $\gamma _{mn}$. It can now be written using the following quadratic formulation as a function of $\zeta$ and $\theta$.
\begin{equation}\label{eq32}
 \gamma _{mn}\left( \zeta, \theta \right) = {\bf{x}}_{mn}^{\rm T}  {{\bf R} \left(\theta \right)}^{\rm T} {{\bf A} \left(\zeta \right)}^{\rm T}   {\bf G} {\bf A}\left(\zeta \right) {\bf R}\left(\theta \right)  {\bf{x}}_{mn},
\end{equation}
where ${\mathbf x}_{mn} $ is the real composite vector form of $\xi_{mn} = g d_{mn}$ given by
\begin{equation}
{\mathbf x}_{mn} = \left[{\xi_{mn}}_{\rm I}  \quad {\xi_{mn}}_{\rm Q}  \right]^T,
\end{equation}
and ${\M{G}}$ contains the statistical characteristics of the aggregate noise including in-phase noise variance, quadrature-phase noise variance, and the correlation between these components. 
\begin{equation}
{\M{G}} =\left[ {\begin{array}{*{20}{c}}
\frac{1}{\sigma_I^2} &  \frac{-\rho_z}{\sigma_I \sigma_Q}\\
\frac{-\rho_z}{\sigma_I \sigma_Q}&\frac{1}{\sigma_Q^2} 
\end{array}} \right].
\end{equation}
Thus, the \ac{BER}  of \ac{HS} can be upper bounded   as
\begin{multline}\label{eq35}
{{\rm{P}}_{\rm b,HS}^{\rm UB}}  \left( {\mathbf{p}},\zeta,\theta \right)  = \frac{1}{{{{\log }_2} \left( M_{\rm nu} \right)}}\sum\limits_{m = 1}^{M_{\rm nu}} {\sum\limits_{{n = 1}\atop
{n \ne m}}^{M_{\rm nu}} {{p_m}} } \\
 \times \mathcal{Q} \left( 
\frac{\sqrt{1 - {\rho _z^2}}}{\sqrt{\alpha \gamma _{mn}\left( \zeta, \theta \right) }}  \ln \left( {\frac{{{p_m}}}{{{p_n}}}} \right) + \frac{\sqrt{\alpha \gamma _{mn}\left( \zeta, \theta \right) }}{2\sqrt{1 - {\rho _z^2}} } \right). 
\end{multline}
\subsection{Problem Formulation}
\ac{HS} targets the joint design of \ac{PS} PMF ${\mathbf{p}}$ and \ac{GS} parameters involving translation $\zeta$ and rotation $\theta$ parameter to minimize the \ac{BER}  bound given in \eqref{eq35}. 
\begin{subequations}\label{eq.P2}
\begin{alignat}{2}
\textbf{P2}:\quad &\!\!\!\!\underset{{{{\bf p}\in \mathbb{S},}  { 0 \le \zeta  \le 1,}} \atop {0 \le \theta  \le 2{\pi}}  }{\text{minimize }}
&&{{\rm{P}}_{\rm b,HS}^{\rm UB}}\left( {\mathbf{p},\zeta,\theta}  \right) \\
&\!\!\!\! \text{subject to} \quad
& & \sum\limits_{m = 1}^{M_{\rm nu}} {{{\left| {{v_m}} \right|}^2}{p_m} \le 1}, \label{eq.2pc} \\
&&& {\rm H}({\mathbf{p}})  \geq \log_2\left({M_{\rm u}}\right), \label{eq.2rc}
\end{alignat}
\end{subequations}
where the average power constraint \eqref{eq.pc} is updated as 
\eqref{eq.2pc} to account for the possible change in the power of the symbols by geometrically shaping the constellation. However, the proposed rate constraint \eqref{eq.2rc} remains intact. Additionally, there are some boundary constraints on $\zeta$ and $\theta$, respectively. 

Intuitively, it is quite difficult to tackle this non-convex multimodal joint optimization problem. Therefore, we resort to the alternate optimization of \ac{PS} parameters (${\mathbf{p}}$) and \ac{GS} parameters ($\zeta,\theta$) using sub-problems $\textbf{P2a}$ and $\textbf{P2b}$, respectively.
Problem $\textbf{P2a}$ designs the PS parameters for some given $\zeta$ and $\theta$. It is quite similar to the problem $\textbf{P1}$ and thus, can be solved using Algorithm \ref{Algo:SCP}.
\begin{align}\label{eq.P2a}
\textbf{P2a}: \quad &\!\!\!\!\underset{{{\bf p}\in \mathbb{S}}}{\text{minimize }}
&& \!\!\!\! {{\rm{P}}_{\rm b,HS}^{\rm UB}}\left( {\mathbf{p},\zeta,\theta}  \right)  \\
&\!\!\!\! \text{subject to} 
& & \!\!\!\! \eqref{eq.2pc}, \eqref{eq.2rc}. \nonumber
\end{align}
On the other hand, the \ac{GS} optimization problem designs $\zeta$ and $\theta$ for fixed symbol probabilities $\bf{p}$, given as
\begin{align}\label{eq.P2b}
\textbf{P2b}:\quad &\!\!\!\!\underset{  { 0 \le \zeta  \le 1,} \atop {0 \le \theta  \le 2{\pi}}  }{\text{minimize }}
{{\rm{P}}_{\rm b,HS}^{\rm UB}}\left( {\mathbf{p}},\zeta,\theta \right) .
\end{align}
The optimization problem \textbf{P2b} is a multimodal non-convex problem which is hard to tackled even by the \ac{SCP} approach as employed in Section \ref{ssec:PS}. The difficulty arises due to the absence of any constraints which restrict the feasibility region. The feasibility space enclosed by the boundary constraints is highly insufficient to serve our purpose.
Therefore, we can approximate the solution using any of the following two methods
\begin{itemize}
\item Trust region reflective method: This method defines a trust region around a specific initial point and then approximate the function within that region. The convex approximation is the first order Taylor series approximation using the gradient.
It begins by
minimizing  convex approximation of the function to obtain a solution.    
This solution is the perturbation in the initial point rendering a new point which should minimize the original function. Otherwise, we need to shrunk the trust region and repeat the process. Reflections are used to increase the step size while satisfying box constraints. After each iteration, we receive a new point which renders a lower objective function than the initial point. This iterative approach leads us to a local minimum and stops when some specified stopping criterion are met \cite{more1983computing,byrd1988approximate}.
\item Gradient descent: This method is a relatively faster approach to tackle the problem at hand. It is owing to the fact that it does not involve any approximation and underlying optimization. It begins with an initial point and keeps updating the point in the descent direction using the gradients and a step size until it reaches a local solution or satisfies some stopping criterion \cite{boyd2004convex}.
\end{itemize}
Interestingly, both of these methods require the gradients of ${{\rm{P}}_{\rm b,HS}^{\rm UB}}\left( {\mathbf{p}},\zeta,\theta \right)$ with respect to $\zeta$ and $\theta$. Gradients are used either to approximate the function with it's first order Taylor series approximation within a trust region or to find the next point in the  descent direction. The gradients are evaluated and presented in Appendix \ref{AppendD}. 
    
    \begin{figure*}
        \centering
        \begin{subfigure}[b]{0.4\textwidth}
            \centering 
       \includegraphics[width=\textwidth]{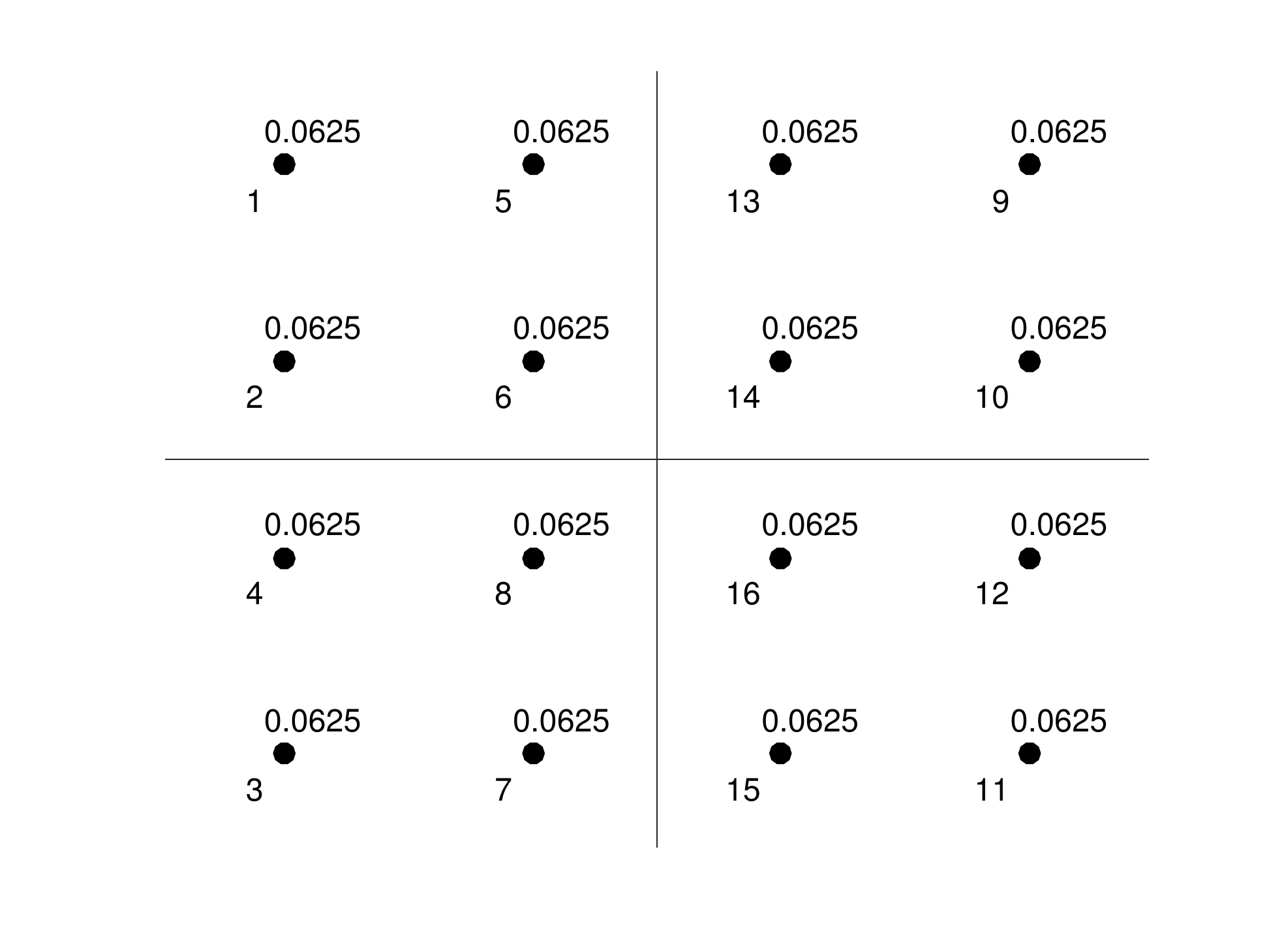} \label{fig:3a}
          \end{subfigure}
   \qquad
        \begin{subfigure}[b]{0.4\textwidth}      \centering 
          \includegraphics[width=\textwidth]{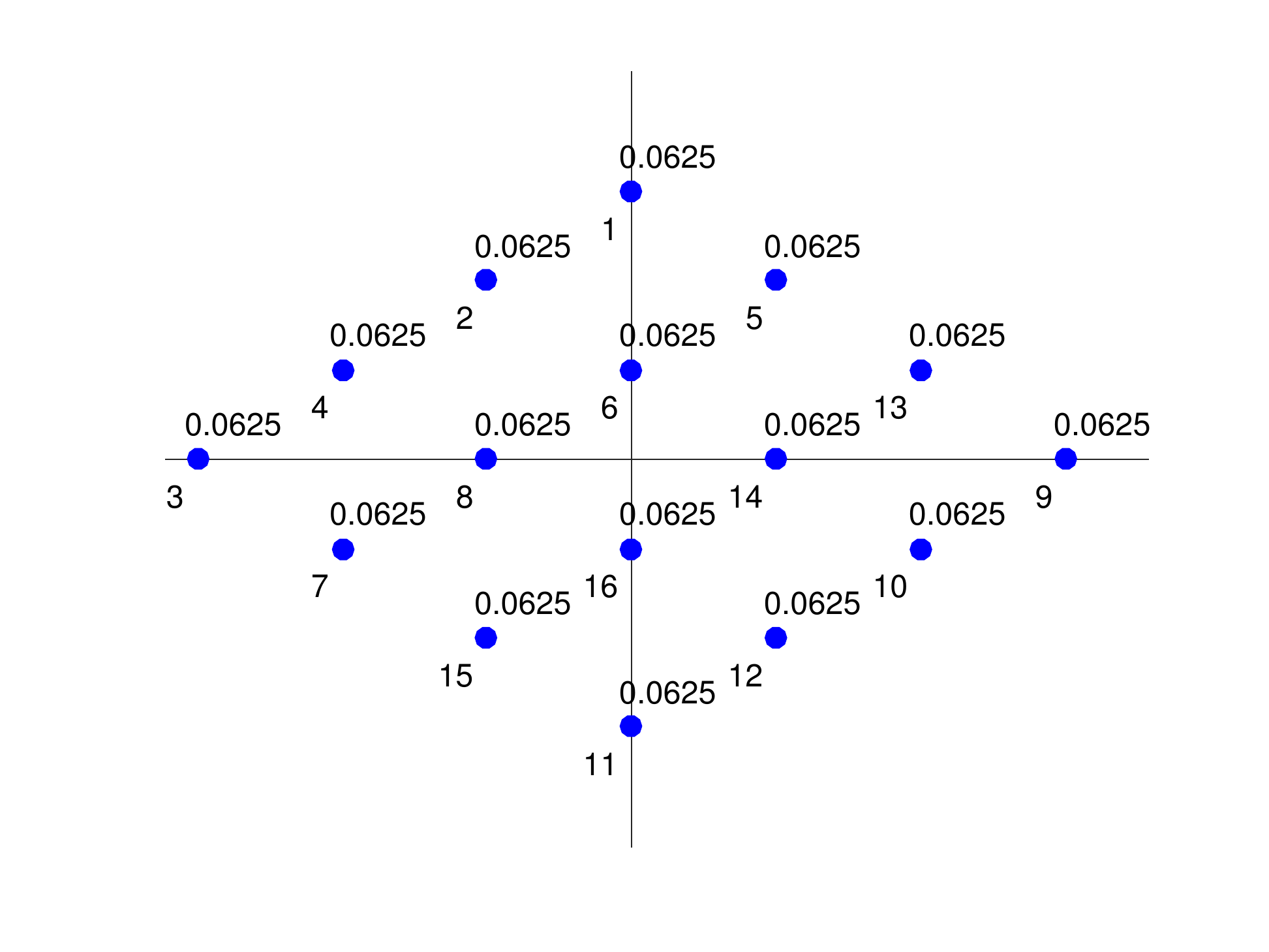} 
            \label{fig:3b}
        \end{subfigure}\\
          {{\small  (a) No Shaping}} 
          \qquad \qquad \qquad  \qquad \qquad \qquad  \qquad \qquad
         {{\small (b) Geometric Shaping}} 
             \vskip\baselineskip
       \begin{subfigure}[b]{0.4\textwidth}
            \centering   \label{fig:3c}
          \includegraphics[width=\textwidth]{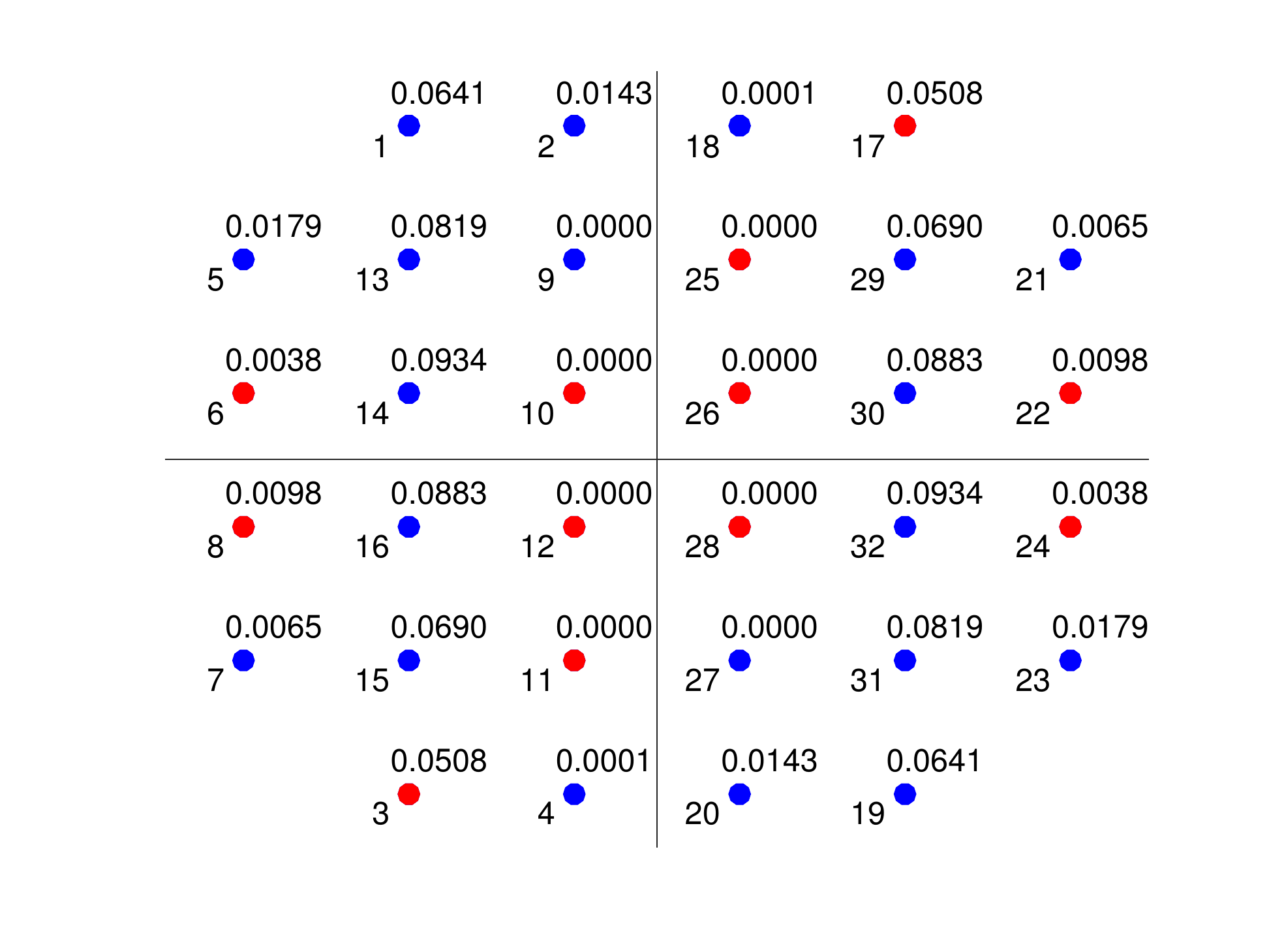}
        \end{subfigure}
        \qquad
        \begin{subfigure}[b]{0.4\textwidth}    \centering   \label{fig:3d}
          \includegraphics[width=\textwidth]{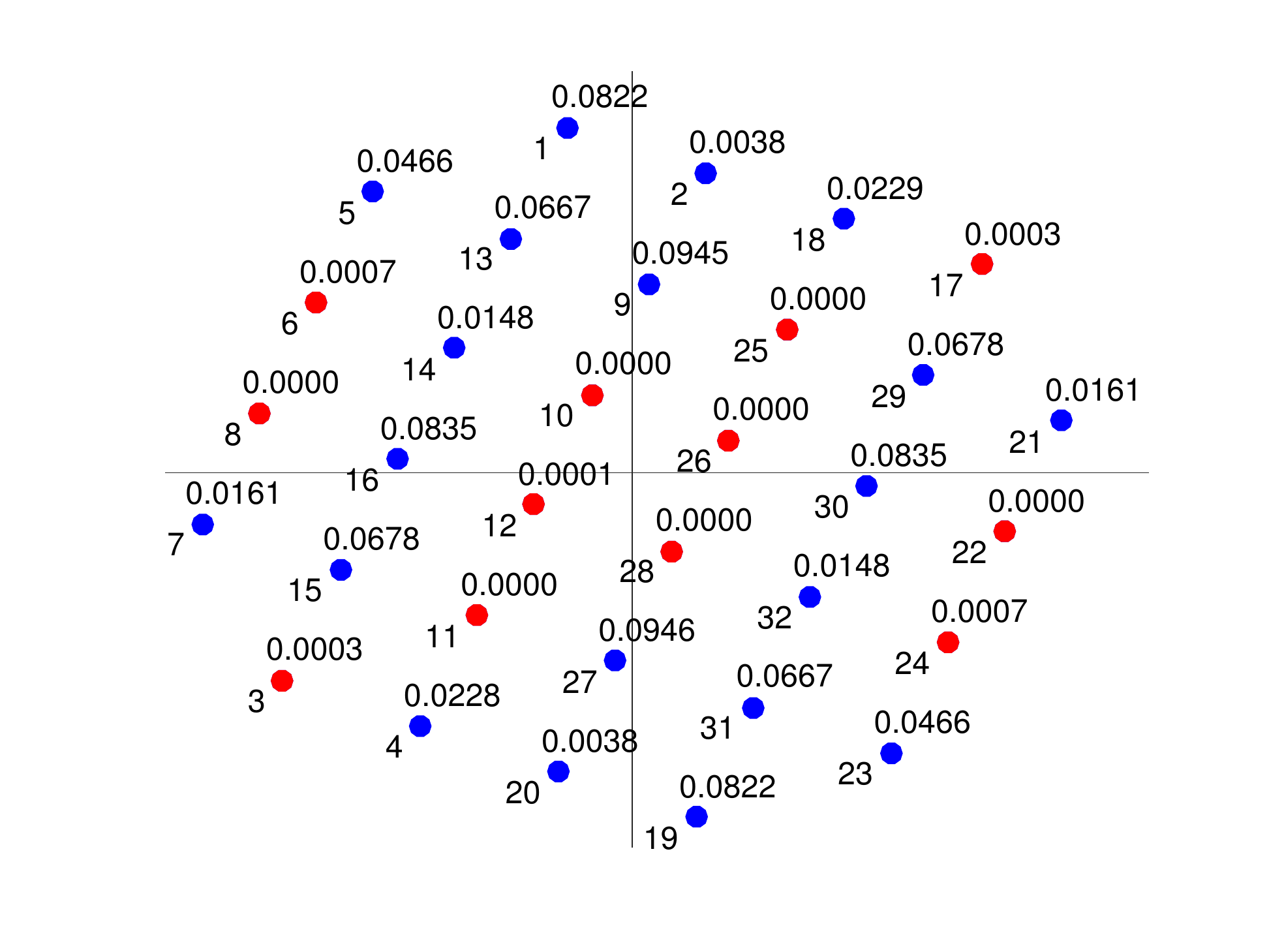} 
        \end{subfigure}\\
        {{\small  (c) Probabilistic Shaping}}
        \qquad \qquad \qquad \qquad \qquad \qquad  \qquad  \quad 
       {{\small  (d) Hybrid Shaping }} 
        \caption[ ]
        {\small Different Asymmetric Signaling Designs} 
        \label{fig:AS}
    \end{figure*}

\subsection{Proposed Algorithm}
The joint optimization problem \textbf{P2} can be tackled using the 
alternate optimization algorithm as presented in Algorithm \ref{Algo:Hybrid}. It solves the sub problems \textbf{P2a} and \textbf{P2b} alternately and iteratively. It begins with some starting feasible points $ {\mathbf{p}}^{(j)}$, $\zeta^{(j)}$, and $\theta^{(j)}$ and evaluates $ {{\rm{P}}_{\rm b,HS}^{{\rm UB}{(j)}}}\left( {\mathbf{p}}^{(j)},\zeta^{(j)},\theta^{(j)} \right)$ as a benchmark. The alternate optimization begins by solving \textbf{P2a} to minimize
${\rm{P}}_{\rm b,HS}^{\rm UB}$ with respect to ${\mathbf{p}}$ given a pair of $\zeta$ and $\theta$.  It is achieved by replacing all entries of $x_m$ with $v_m = {\bf AR}x_m \; \forall m$.  ${\mathbf{p}}^{(j*)}$ is obtained using the framework provided in Algorithm \ref{Algo:SCP} which solves \textbf{P1a} iteratively. Then, the optimum ${\mathbf{p}}^{(j*)}$ is used as a given PMF to obtain the pair $\zeta^{(j*)}$ and $\theta^{(j*)}$ by solving \textbf{P2b}. These optimum parameter values are updated to attain next initial points. Moreover, $ {{\rm{P}}_{\rm b,HS}^{{\rm UB}{(j+1)}}}\left( {\mathbf{p}}^{(j+1)},\zeta^{(j+1)},\theta^{(j+1)} \right)$ is also evaluated to compare the decrease in objective function. The norm of this difference is stored in $\epsilon$ and the process is repeated until this value drops below a preset threshold $\delta$. Eventually, the solution parameters are updated in $\left({\mathbf{p}}^*, \zeta^*, \theta^* \right)$ which yield the minimized \ac{BER} upper bound ${\rm{P}}_{\rm b,HS}^{{\rm UB}*}$ using \ac{HS}. Therefore, these HS parameters are capable of rendering a BER ${\rm{P}}_{\rm b,HS}^*$ lower than the bound ${\rm{P}}_{\rm b,HS}^{{\rm UB}*}$.
\begin{algorithm}[t!]
\caption{Alternate Optimization}\label{Algo:Hybrid}
\begin{algorithmic}[1]
\State \textbf{Initialize}  $j \gets 0$, $\epsilon \gets \infty$   and   \textbf {Set}   tolerance $\delta$
\State \textbf{Choose} feasible starting points $ {\mathbf{p}}^{(j)}$, $\zeta^{(j)}$, and $\theta^{(j)}$.
\State \textbf{Evaluate} $ {{\rm{P}}_{\rm b,HS}^{{\rm UB}{(j)}}}\left( {\mathbf{p}}^{(j)},\zeta^{(j)},\theta^{(j)} \right)$.
\While {$\epsilon \ge \delta$}
\State Solve \textbf{P2a} using Algorithm \ref{Algo:SCP} with starting point $ {\mathbf{p}}^{(j)}$ and given $\zeta^{(j)}$, $\theta^{(j)}$ to obtain $ {\mathbf{p}}^{(j*)}$
\State Solve \textbf{P2b}  with starting points $\zeta^{(j)}$,$\theta^{(j)}$ and given ${\mathbf{p}}^{(j*)}$ to obtain $\zeta^{(j*)}$, $\theta^{(j*)}$
\State   ${{\mathbf{p}}}^{(j+1)} \gets {\mathbf{p}}^{(j*)}$, $\zeta^{(j+1)} \gets \zeta^{(j*)}$, and $\theta^{(j+1)} \gets \theta^{(j*)}$   
\State Evaluate $ {{\rm{P}}_{\rm b,HS}^{{\rm UB}{(j+1)}}}\left( {\mathbf{p}}^{(j+1)},\zeta^{(j+1)},\theta^{(j+1)} \right)$.
\State Update ${\epsilon \gets \left\| {{\rm{P}}_{\rm b,HS}^{{\rm UB}{(j+1)}}}- {{\rm{P}}_{\rm b,HS}^{{\rm UB}{(j)}}} \right\|}$
\State   ${j \gets j+1}$
\EndWhile
\State  Solution parameters: ${\mathbf{p}}^* \gets {\mathbf{p}}^{j+1}$, $\zeta^* \gets \zeta^{j+1}$, $\theta^* \gets \theta^{j+1}$
\State   Objective function: ${{\rm{P}}_{\rm b,HS}^{{\rm UB}{*}}} \gets {{\rm{P}}_{\rm b,HS}^{{\rm UB}{(j+1)}}}$
\State  Consequence: $ \rm{P}_{\rm b,HS}^* \leq {{\rm{P}}_{\rm b,HS}^{{\rm UB}{*}}}$
\end{algorithmic}
\end{algorithm}

Numerical evaluations reveal that the stopping criteria is mostly met in just one iteration. Interestingly, Step 5 and 6 in Algorithm \ref{Algo:Hybrid} are interchangeable and need to be chosen carefully. For instance, \ac{PS} demonstrates better performance at higher  \ac{HWD} levels so it is intuitive to design the \ac{HS}  by first \ac{PS} and then \ac{GS} in order to attain further gain over PS. Whereas, \ac{GS} depicts lower \ac{BER}  at lower  \ac{HWD} levels so it is recommended to design \ac{HS}  by first \ac{GS} and then \ac{PS} in order to achieve better performance than \ac{GS} using the added DoF offered by PS.

\ac{HS}  can be implemented by choosing the transmit symbols for the translated and rotated signal constellation, i.e., $ v_m=\M{A} \left(\zeta^*\right) \M{R}\left( \theta^*\right) x_m$. Furthermore, the symbols are transmitted according to the optimized  ${\mathbf{p}}^*$ where $\zeta^*$, $\theta^*$ and ${\mathbf{p}}^*$ are designed using Algorithm \ref{Algo:Hybrid}. Upon reception, they are detected using the \ac{MAP}  detector as presented in \eqref{eq30}.
\subsection{Illustrative Example}
We present a comprehensive example to highlight the design of various distinct shapes for a fixed rate of 4bits/symbol. The black color is used for the reference constellation. The blue color depicts the possible transmission symbols whereas red symbols highlight the improbable transmission symbols. Fig. 3a presents uniformly distributed $16$-QAM constellation with no-shaping. Fig. 3b illustrates geometrically shaped $16$-QAM with parameters $\zeta=0.5$ and $\theta=\pi/2$. The parallelogram envelop encloses equally prior QAM symbols.  Next, we employ $32$-QAM and design non-uniform probabilities as detailed in section IV.  The red symbols highlight the symbols with negligible transmission probabilities whereas blue symbols have some notable transmission probabilities as depicted in Fig. 3c. The proposed algorithm tends to discard symbols with minimal transmission power to reduce the \ac{BER}. One possible reason is that these symbols are mostly affected by the improper  \ac{HWD} owing to their comparable power/variance. Furthermore, this probabilistic shaped constellation undergoes \ac{GS} to demonstrate hybrid shaped QAM constellation as shown in Fig. 3d. 

\section{Numerical Results}\label{sec:Numerical}
Numerical evaluations of the adopted \ac{HWD} system are carried out to study the drastic effects of hardware imperfections and the effectiveness of the mitigation strategies. The performance of the proposed \ac{PS}  and \ac{HS}  as a realization of asymmetric transmission is quantified with varying energy per bit per noise ratio (EbNo) and  \ac{HWD} levels. EbNo is obtained by normalizing SNR with the transmission rate.  The derived error probability bounds and performance of the asymmetric transmission schemes are also validated using Monte-Carlo simulations. We compare the performance of conventional \ac{GS} with the proposed PS. \ac{GS} can be implemented by transmitting symbols from a reshaped constellation $\V{v}_m = \M{A}\left( \zeta^*\right)\M{R}\left( \theta^*\right)\V{x}_m$, where $\zeta^*$ and $\theta^*$ can be obtained by solving \textbf{P2a} given uniform prior distribution. Upon reception, they are detected using the ML detector which is the simplified form of optimal \ac{MAP}  detector \eqref{eq30} given uniform prior probabilities. This ML detector considers the reshaped constellation symbols $\V{v}_m$ as the reference to detect the received symbols. 

For most of the numerical evaluations we  assume grey coded square QAM constellations of order $M_{\rm u}=8$, i.e., $R= \log_{2}(M_{\rm u})$, for no-shaping (NS) and \ac{GS} as benchmarks. For \ac{PS} and \ac{HS}  we employ  $M_{\rm nu}=32$-QAM with rate at least as high as that of GS, i.e., $R\geq \log_{2}(M_{\rm u})$. Moreover, we consider practical  \ac{HWD} values for the transmitter $\kappa_t = 0.01$ and receiver $\kappa_r = 0.12$. The pseudo-variances are derived from the   $\tilde{\kappa}_{\rm tI} = \kappa_t /4$,  $\tilde{\kappa}_{\rm rI} = \kappa_r /4$,  and correlation coefficient $\rho_\eta = 0.9$. Intuitively, AWGN channel assumes $g=1$ and circularly symmetric Rayleigh fading channel is generated using $\lambda=1$. Furthermore, the transmission EbNo is taken as $30$~dB. The aforementioned values of the parameters are used throughout the numerical results, unless specified  otherwise.

\begin{figure}[t]
    \centering
   \includegraphics[width=3.5in]{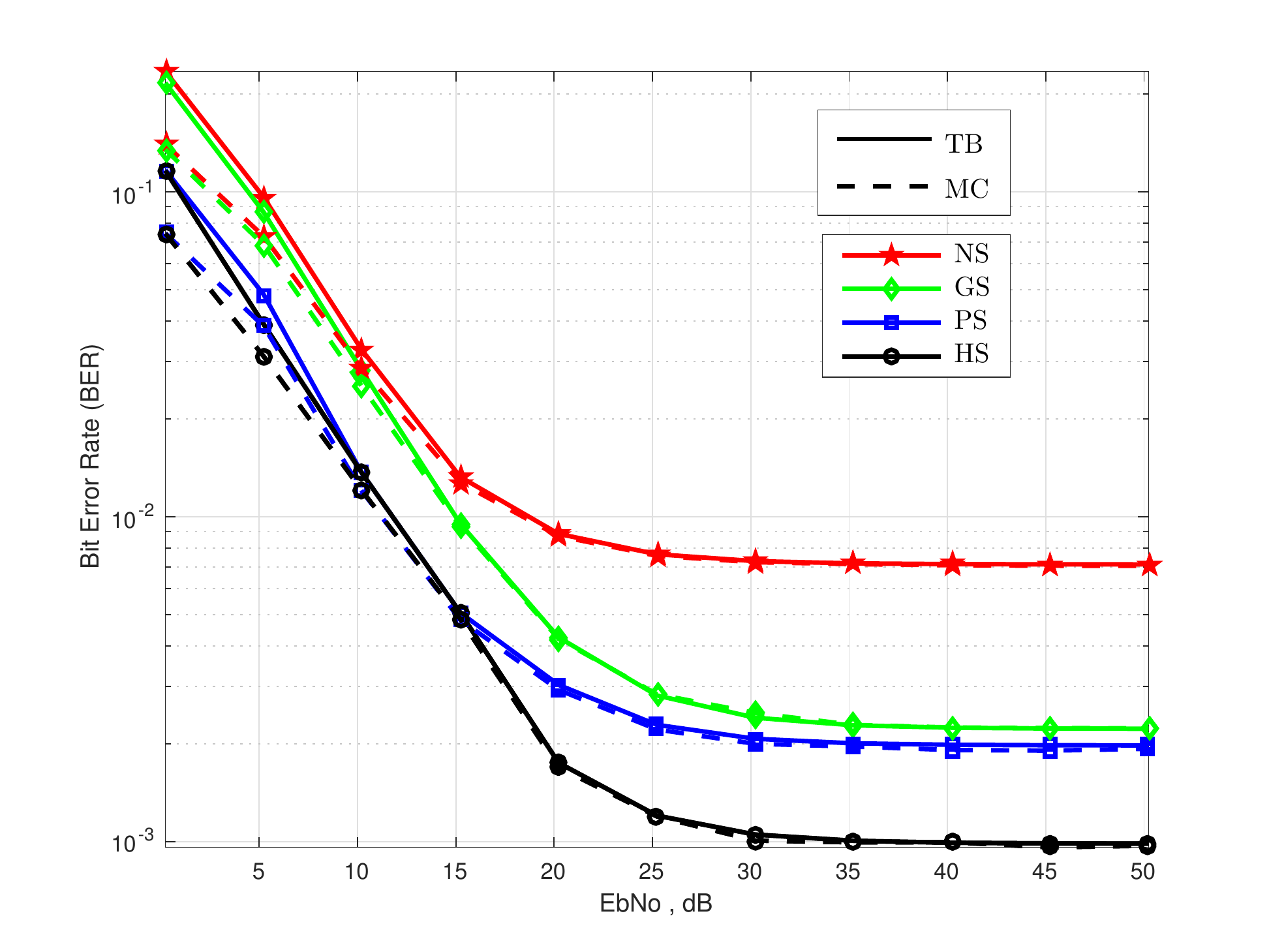}
    \caption{BER performance for a range of EbNo with $\kappa=0.13$ in AWGN channel.}
    \label{fig:BERvsEbNo}
\end{figure}
First, we evaluate the performance of various AS schemes for a range of EbNo from $0$~dB to $50$~dB in an AWGN channel as shown in Fig.~\ref{fig:BERvsEbNo}. We employ $M_{\rm u}$-QAM for NS and \ac{GS} whereas  $M_{\rm nu}$-QAM for \ac{PS} and HS. The \ac{BER} performance improves with increasing EbNo till $30$~dB and then undergoes saturation owing to the presence of  \ac{HWD}. Further increase in bit energy also results in an increase in the distortion variance, as the system experiences an error floor which can be deduced from \eqref{eqAsymp}. Evidently, the proper/symmetric QAM is suboptimal and the \ac{BER}  performance is significantly improved using AS. Conventional \ac{GS} is not beneficial at lower EbNo values, but it significantly improves the performance for higher EbNo values pertaining to the increased symbol space \cite{javed2019asymmetric}.
On the other hand, the proposed \ac{PS} is capable of minimizing the \ac{BER}  for the entire range of EbNo. Substantial gains can be achieved by taking another step forward and employing HS. Therefore, we can safely conclude that the best performance can be achieved using \ac{PS} for  EbNo $\leq$ $15$~dB and \ac{HS}  for EbNo $\geq$ $15$~dB as validated by the Monte Carlo simulation.
%
At $20$~dB, the \ac{BER} reductions for GS, PS, and HS schemes with respect to unshaped constellation are approximately $52.22$\%,  $66.67$\%,  $80$\%, respectively. 

\begin{figure}[t]
    \centering
   \includegraphics[width=3.5in]{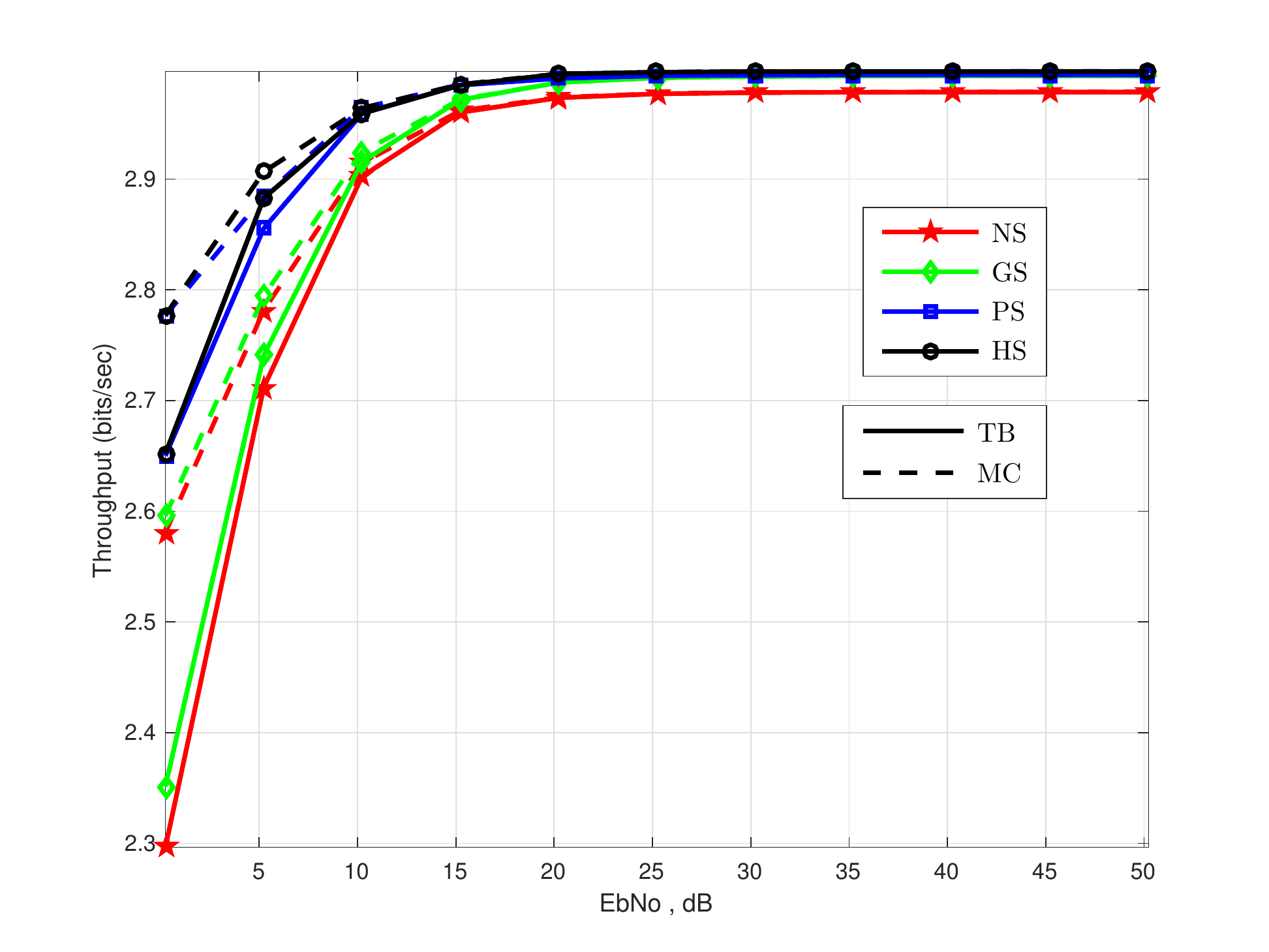}
    \caption{ Throughput performance for a range of EbNo with $\kappa=0.13$ in AWGN channel.}
    \label{fig:ThvsEbNo}
\end{figure}

For the same simulation settings, we analyze system throughput (correct bits/symbol) for a range of EbNo values where the lower bound on system throughput can be obtained as
\begin{equation}
\mathcal{T}^{\rm LB}\left( \mathbf{p} \right) = \left[ 1- {{\rm{P}}_b^{\rm UB}}\left( \mathbf{p}  \right) \right] {\rm{H}}(\mathbf{p} )
\end{equation}
Fig. \ref{fig:ThvsEbNo} depicts negligible throughput gain of \ac{GS} over NS but noticeable throughput improvement using \ac{PS} or HS. For instance,
$1.5$\%, $6$\% and $7$\% percentage increase in throughput can be observed using GS, PS, and \ac{HS}  at $\text{EbNo}=5$~dB.
The throughput gain is quite substantial for lower EbNo values but undergoes saturation when EbNo $\geq$ $20$~dB. Interestingly, PS/HS saturates at $3$~bits/symbol following rate fairness constraint with negligible \ac{BER}  whereas other schemes saturate below $3$~bits/symbol depicting significant \ac{BER}  even though the entropy of $8$-QAM with uniform distribution is $\log_2(8)=3$.

\begin{figure}[htbp]
    \centering
   \includegraphics[width=3.5in]{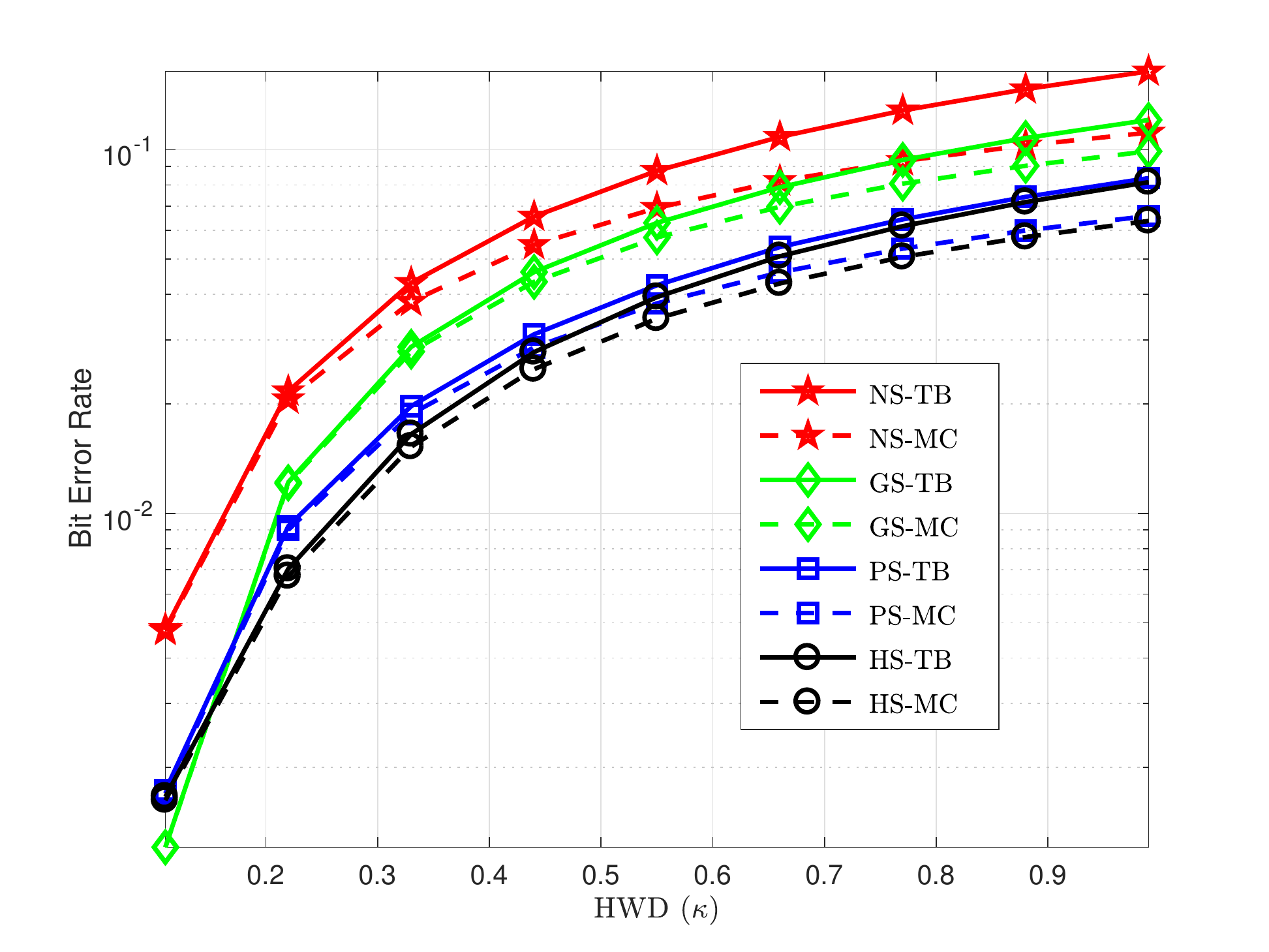}
    \caption{HWD mitigation using $16$-QAM for \ac{PS} and \ac{HS}  at $\text{EbNo}=30$~dB in an AWGN channel.}
    \label{fig:BERvsHWD}
\end{figure}
\begin{figure}[t]
    \centering
   \includegraphics[width=3.5in]{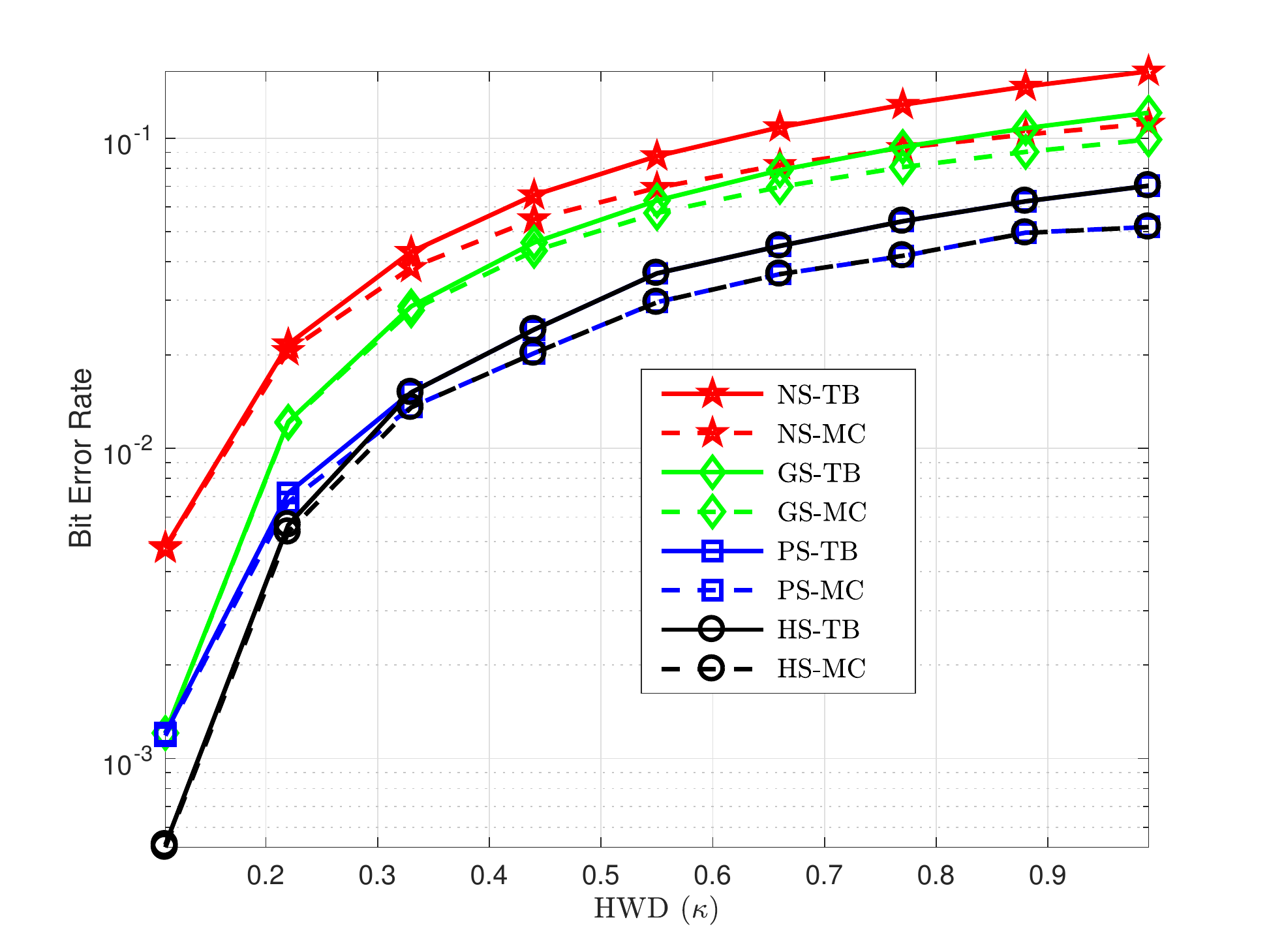}
   \caption{ \ac{HWD} mitigation using $32$-QAM for \ac{PS} and \ac{HS} at $\text{EbNo}=30$~dB in an AWGN channel.}
    \label{fig:BERvsHWD2}
\end{figure}
Next we analyze the behavior of various AS schemes with increasing distortion levels as depicted in Fig. \ref{fig:BERvsHWD}. We assume $8$-QAM for benchmark NS and traditional \ac{GS} whereas $16$-QAM for \ac{PS} and HS. Derived bounds are in close accordance with the MC simulation especially for lower  \ac{HWD} levels. 
Obviously, the \ac{BER}  increases with increasing  \ac{HWD} levels and AS based systems achieve lower \ac{BER}  by efficiently mitigating the drastic  \ac{HWD} effects. Undoubtedly, the NS scheme suffers the most, but \ac{GS} helps to decrease the \ac{BER}  to some extent. Further compensation can be achieved using the proposed \ac{PS} and HS. Surprisingly, \ac{GS} outperforms \ac{PS} and \ac{HS}  at the lowest  \ac{HWD} values, e.g., $\kappa = 0.11$, in Fig.~\ref{fig:BERvsHWD}
but PS/HS maintain their superiority for $\kappa \geq 0.17$.
Interestingly, PS/HS are still capable of outperforming GS
even for the lowest  \ac{HWD} levels pertaining to their rate adaptation capability and added DoF using $32$-QAM as highlighted in Fig. \ref{fig:BERvsHWD2}. We can observe enhanced mitigation offered by the $32$-QAM PS/HS as compared to the $16$-QAM PS/HS due to the added DoF. For instance, we observe \ac{BER}  compensation of $66$\% and $77.5$\% using $32$-QAM \ac{PS} and HS, respectively, whereas \ac{BER}  compensation of $55$\% and $65$\% using $16$-QAM \ac{PS} and HS, respectively, at $\kappa = 0.22$  \ac{HWD} level. 

\begin{figure}[t]
    \centering
   \includegraphics[width=3.5in]{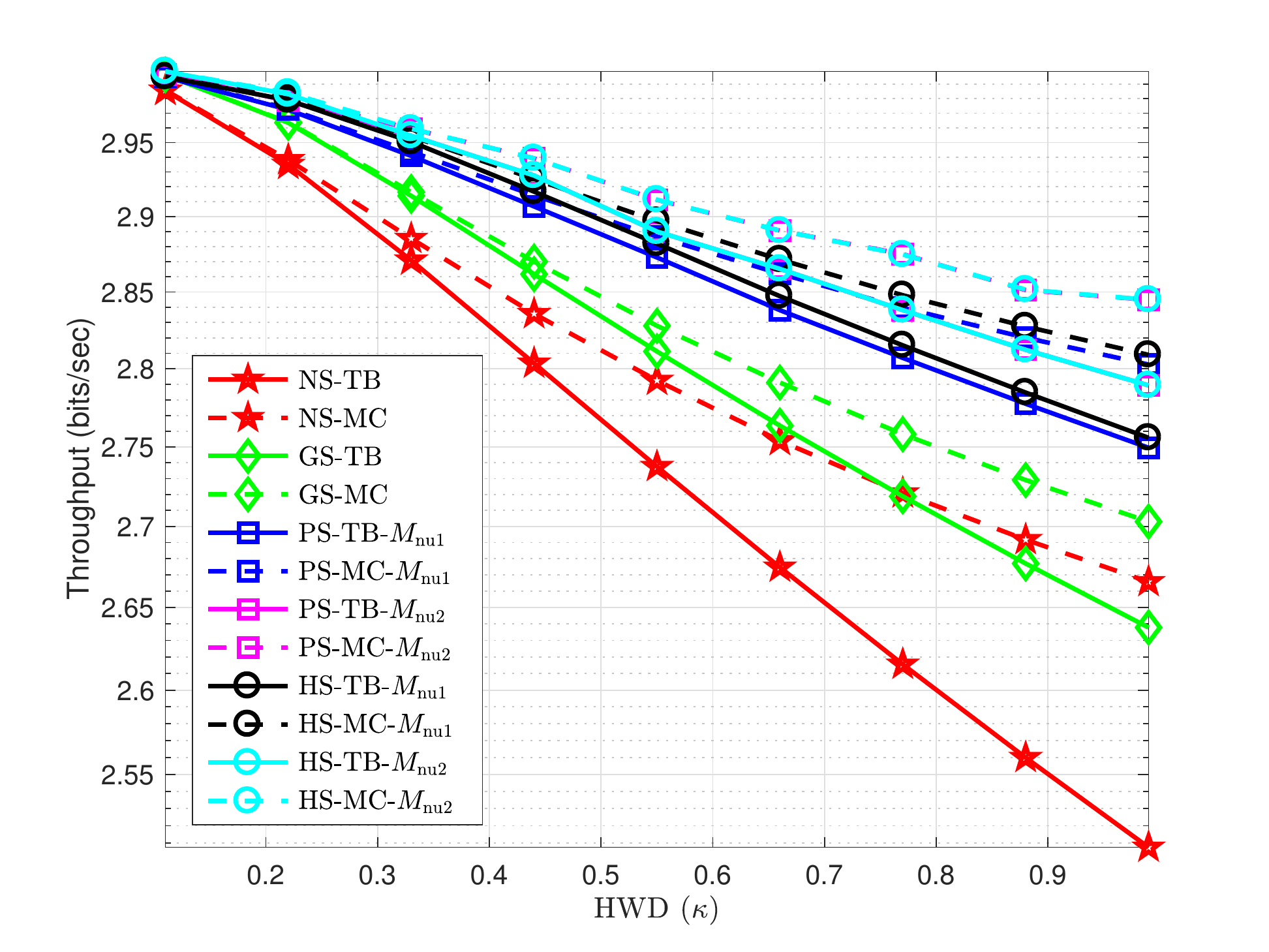}
    \caption{System throughput for a range of  \ac{HWD} levels at $\text{EbNo}=30$~dB in an AWGN channel.}
    \label{fig:ThvsHWD}
\end{figure}

A similar analysis is undertaken to study the impact of increasing  \ac{HWD} on the system throughput. Fig. \ref{fig:ThvsHWD} compares the throughput performance of $M_{\rm u}$-QAM NS and \ac{GS} with $M_{\rm nu1}=16$-QAM \ac{PS} and \ac{HS}  as well as with $M_{\rm nu2}=32$-QAM \ac{PS} and HS. System throughput decreases almost linearly with increasing  \ac{HWD} for all forms of signaling but with different slopes. NS demonstrates the steepest slope with increasing  \ac{HWD} and all the other AS schemes render gradual slopes. Quantitative analysis shows the slopes of $-0.55$, $-0.41$, $-0.28$, and  $-0.24$ using NS, GS, $16$-QAM PS/HS, and $32$-QAM PS/HS, respectively, with increasing  \ac{HWD}. Therefore, \ac{PS} and \ac{HS}  present the most favorable results as compared to the GS. Their performance can be even improved by increasing the modulation order. Another important observation is the 
overlapping response of \ac{PS} and \ac{HS}  especially for higher ordered QAM, which suffices \ac{PS} and revokes the need of \ac{HS}  to perform even better.

\begin{figure}[t]
    \centering
   \includegraphics[width=3.5in]{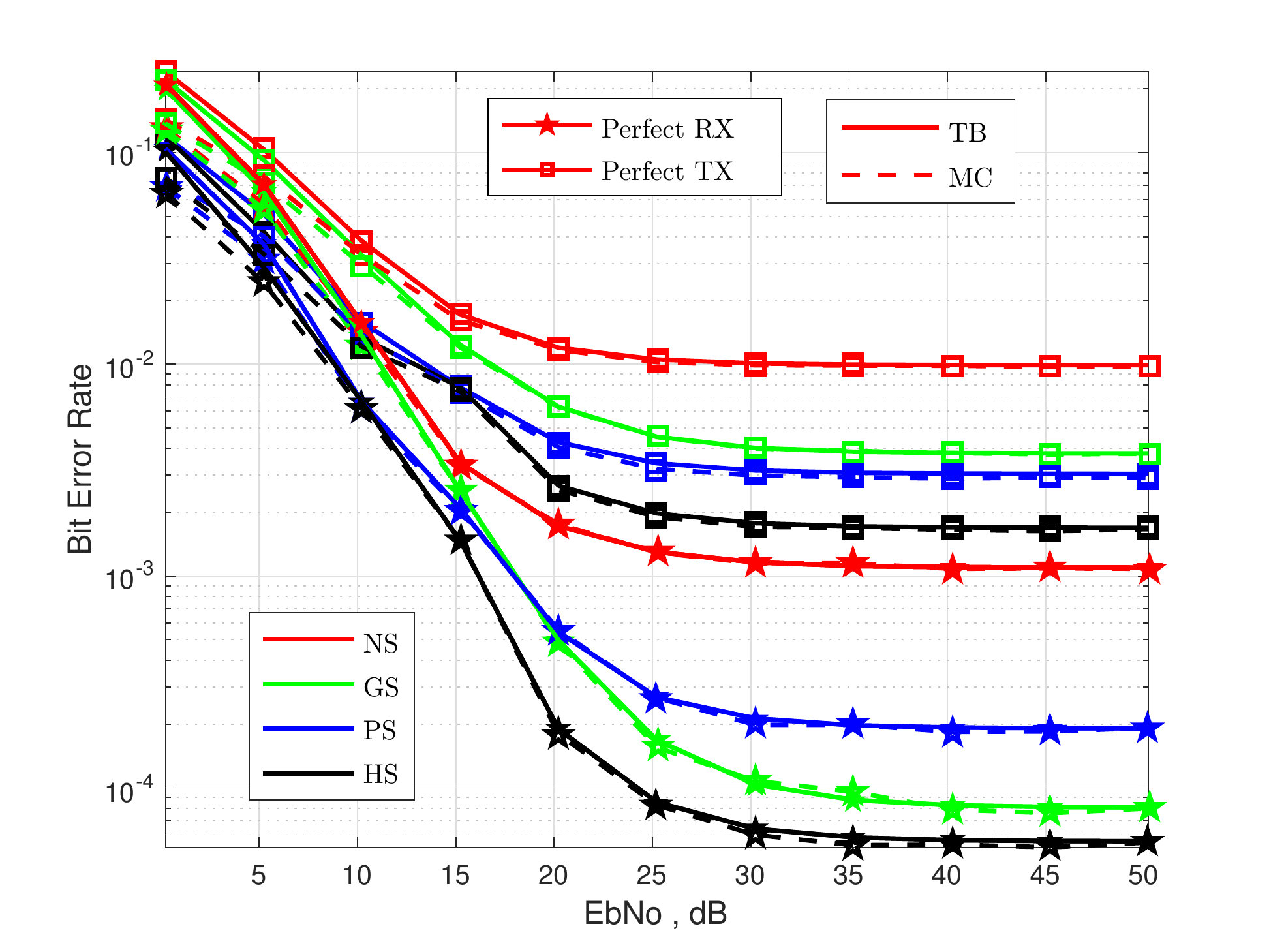}
    \caption{BER performance for two different  \ac{HWD} scenarios (Perfect Receiver: $\kappa_t = 0.07$ and Perfect Transmitter: $\kappa_r = 0.15$) in an AWGN channel.}
    \label{fig:BERforTRX}
\end{figure}

Another simulation example depicts the performance of the discussed AS schemes with  over a range of EbNo for two distinct scenarios of perfect receiver and perfect transmitter as presented in Fig. \ref{fig:BERforTRX}. Perfect receiver system as the name specifies includes ideal zero-distortion receiver but imperfect transmitter with  $\kappa_t = 0.07$ whereas perfect transmitter system involves ideal zero-distortion transmitter but imperfect receiver with $\kappa_r = 0.15$.  Note that the lower value of $\kappa_t$ relative to $\kappa_r$ is due to the fact that transmitters employ sensitive equipment to exhibit low distortions because the transmitter distortions are far more drastic than the receiver distortions. Interestingly, \ac{GS} outperforms \ac{PS} at EbNo > $15$~dB for the perfect receiver case as opposed to EbNo < $15$~dB where \ac{PS} is still a better choice. \ac{HS}  outperforms both of them irrespective of the EbNo range classification. At such low  \ac{HWD} level, the \ac{BER}  percentage reduction of  
$81.82$\%,  $90.91$\%,  $94.55$\% is observed using PS, GS, and HS
at $30$~dB EbNo. 
Regarding the perfect transmitter scenario, \ac{GS} and \ac{PS} reverse the trend for higher EbNo level. Now the \ac{PS} clearly outperforms \ac{GS} for the entire range of EbNo and the \ac{HS}  marks its superiority over both of these schemes. At $0.15$  \ac{HWD} level, the EbNo gain of $8$~dB, $12$~dB, and $13$~dB are estimated using GS, PS, and \ac{HS}  to attain the \ac{BER}  of $10^{-2}$. 

\begin{figure}[t]
    \centering
   \includegraphics[width=3.5in]{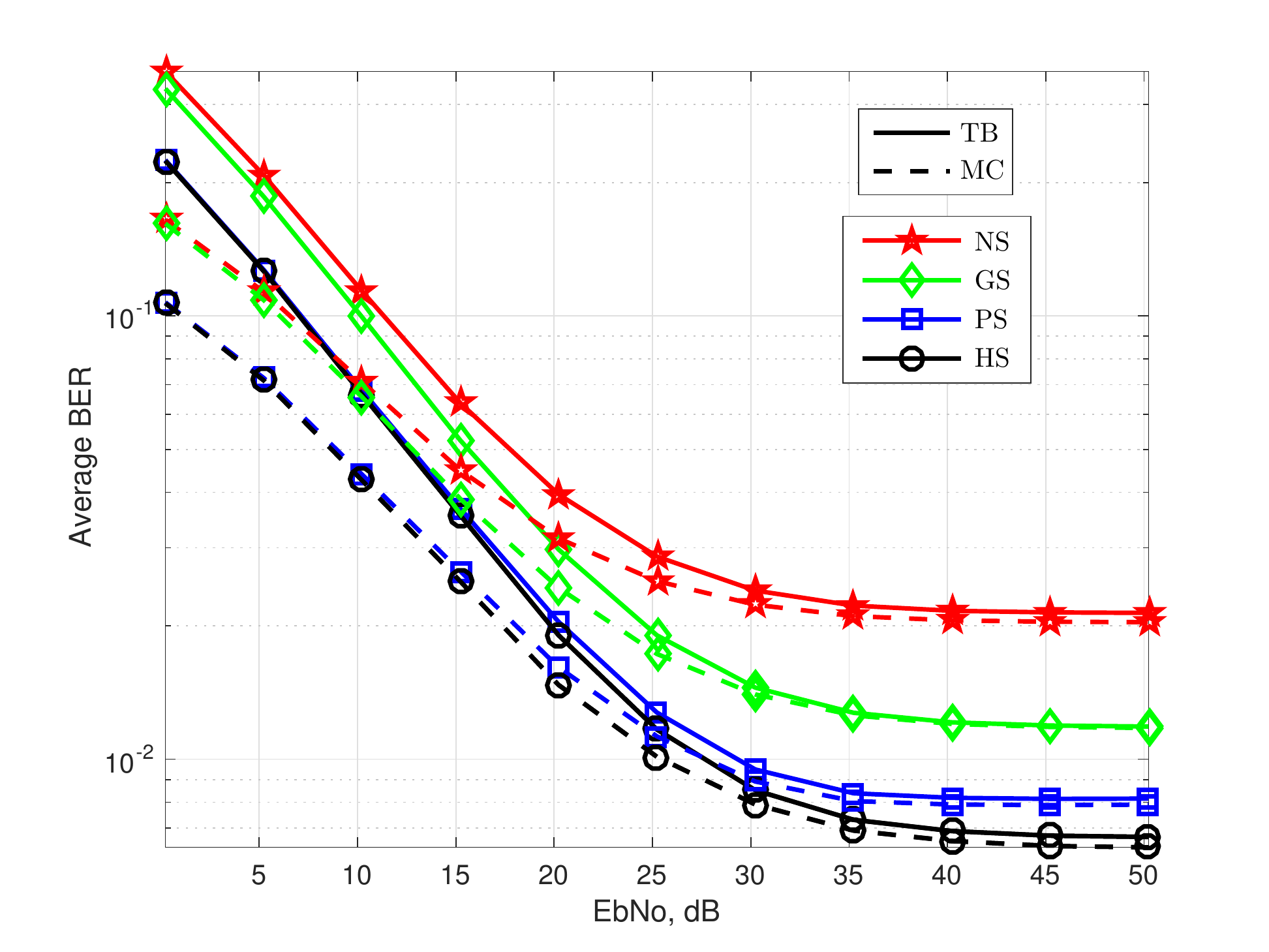}
    \caption{Average \ac{BER} performance with  \ac{HWD} $\kappa = 0.22$ in a Rayleigh fading channel.}
    \label{fig:BER_Rayleigh}
\end{figure}

Finally, the average (ergodic) \ac{BER}  performance of the adopted system with $\kappa=0.22$  \ac{HWD} level is evaluated over a Rayleigh fading channel for a range of EbNo values as given in Fig. \ref{fig:BER_Rayleigh}. Evidently, the \ac{AS}  schemes preserve their \ac{BER}  trends and order.  Clearly, average \ac{BER}  decreases with increasing EbNo and then undergoes saturation yielding an error floor. The derived \ac{BER}  bounds are also validated using MC simulations rendering a tighter bound for higher EbNo values. \ac{GS} improves the average \ac{BER}  as compared to the NS scenario but \ac{PS} and \ac{HS}  maintain their superior performance. Signaling schemes of GS, PS, and \ac{HS}  offer a
percentage reduction of  $54.55$\%, $63.64$\%, and $70.45$\%, respectively,  in the  average \ac{BER}  performance at $40$~dB EbNo.

In a nutshell, we can conclude that the \ac{GS} offers significant \ac{BER}  reduction at higher \ac{SNR}  values as opposed to the \ac{PS} which offers universal gains. Moreover, the perks of \ac{HS}  are also prominent for higher \ac{SNR}  and higher $M$-ary modulation but depicts \ac{PS} comparable performance at lower \ac{SNR}  values. Therefore, we recommend to employ \ac{HS}  given high \ac{SNR}  but resort to \ac{PS} for lower \ac{SNR}  values to save additional computational expense.  Additionally, \ac{GS} is a better choice for slightly distorted systems whereas PS/HS are the optimal choice for moderate to severely distorted systems. Furthermore, we can achieve improved performance by employing higher-order QAM constellations for PS/HS given adequate resources. On the other hand, the throughput gains are eminent at considerably lower \ac{SNR}  values and higher distortion values.

 \begin{table*}[!t]
\caption{First Order Necessary KKT Conditions}
\centering
  \begin{tabular}{|c|c|c|c|}
     \hline
    \hline
Index & KKT Conditions & Satisfied with &Reason \\
      \hline
    \hline
$1:M_{\rm nu}$ & $\nabla_{\mathbf{p}}\mathcal{L} \left(   {\mathbf{p}}^*,  {\lambda ^*}\right)=0, \; \forall 1 \leq m \leq M_{\rm nu}$ & $\nabla_{\mathbf{p}}\mathcal{L} \left(   {\mathbf{p}}^*,  {\lambda _1^*},{\lambda _2^*,{\lambda _3^*}}\right)=0$  & Saddle point of the dual problem \\
    \hline
$M_{\rm nu}$+1 &  ${\lambda _1^*}\left( {\sum\limits_{m = 1}^{M_{\rm nu} } {{\left| {{x_m}} \right|}^2}{p_m^*} - 1} \right) = 0$ &$ {\sum\limits_{m = 1}^{M _{\rm nu}} {{\left| {{x_m}} \right|}^2}{p_m^*} }  = 1$,  ${\lambda _1^*} \geq 0$ & Maximum power transmission \\
     \hline
$M_{\rm nu}$+2 & ${\lambda _2^*}\left( {\sum\limits_{m = 1}^{M_{\rm nu} } {p_m^*} - 1} \right) = 0$ & $ {\sum\limits_{m = 1}^{M_{\rm nu} } {p_m^*} } = 1$, ${\lambda _2^*} \geq 0$ & Equality constraint  \\
    \hline
$M_{\rm nu}$+3 & $ {\lambda _3^*} \left(  \log_2\left({M_{\rm u}}\right)-{\rm H}({\mathbf{p}}^*) \right)=0$ &  $ {\rm H}({\mathbf{p}}^*)  =  \log_2\left({M_{\rm u}}\right)$, ${\lambda _3^*} \geq 0$  & \ac{BER} -Rate tradeoff  \\
    \hline
  \hline
  \end{tabular}
  \label{tab:TableKKT}
\end{table*}

\section{Conclusion}
This work proposes probabilistic and hybrid shaping to realize asymmetric signaling in digital wireless communication systems suffering from improper  \ac{HWD}. Instinctively, all forms of asymmetric shaping are capable of decreasing the \ac{BER}, and this performance gain improves with increasing \ac{SNR}  and/or increasing  \ac{HWD} levels with respect to NS. However, \ac{PS} outperforms \ac{GS} and performs equally well as  HS. We can achieve more than $50$\% \ac{BER}  reduction with PS/HS over traditional GS. 
The perks of \ac{PS} come at the cost of increased complexity in the design and decoding process. The \ac{HS}   scheme is capable of improving the system performance in terms of the \ac{BER}  as well as throughput. However, for less  \ac{HWD} levels and low EbNo, the benefits of \ac{HS}  over \ac{PS} are limited while requiring additional complications in optimization, modulation, and detection procedures. Therefore, \ac{PS} emerges as the best choice in the trade-off between enhanced performance and added complexity. 

\appendices
\section{Statistical Characterization of Aggregate Noise}
\label{AppendA}
The superposed Gaussian distributions render the accumulative noise $z \sim \mathcal{CN} \left( {0,v,\tilde v} \right)$, where $v=\alpha {{\left| g \right|}^2}\kappa  + \sigma _w^2$ and ${\tilde v} = \alpha {g^2}\tilde \kappa$. Exploiting the relation between the $v$, $\tilde v$ and the variances of $\sigma _I^2 = E \{z_I^2\}$, $\sigma _Q^2  = E\{z_Q^2\}$ and their mutual correlation ${r_{{z_I}{z_Q}}} = E\left\{ {{z_I}{z_Q}} \right\}$, we get
\begin{equation}
v = {E\left\{ {{{\left| z \right|}^2}} \right\}  = \sigma _I^2 + \sigma _Q^2 }.
\end{equation}
\begin{equation}
{\tilde v}= E\left\{ {{z^2}} \right\}   = \sigma _I^2 - \sigma _Q^2 + i 2{r_{{z_I}{z_Q}}}.
\end{equation}
Their inter relation enables us to evaluate $\sigma _I^2 $, $\sigma _Q^2$, and ${r_{{z_I}{z_Q}}} $ from $v$ and $\tilde{v}$ as
\begin{align}\label{A4}
&{\sigma _I^2 = \frac{{v + {{\tilde v}_I}}}{2} = \frac{{\alpha {{\left| g \right|}^2}\kappa  + \sigma _w^2 + \alpha {\Re}\left( {{g^2}\tilde \kappa } \right)}}{2}}, \\
& 
{\sigma _Q^2 = \frac{{v - {{\tilde v}_I}}}{2} = \frac{{\alpha {{\left| g \right|}^2}\kappa  + \sigma _w^2 - \alpha {\Re}\left( {{g^2}\tilde \kappa } \right)}}{2}}, \label{A5} \\
& 
{r_{{z_I}{z_Q}}} =\frac{{{{\tilde v}_Q}}}{2} = \frac{{\alpha {{\Im}}\left( {{g^2}\tilde \kappa } \right)}}{2}. \label{A6}
\end{align}
Finally, \eqref{A4}-\eqref{A6} allow us to find the correlation coefficient between $z_I$ and $z_Q$ as
\begin{equation}\label{A7}
\rho_z  = \frac{{{r_{{z_I}{z_Q}}}}}{{{\sigma _I}{\sigma _Q}}} = 
{\frac{{\alpha {\Im}\left( {{g^2}\tilde \kappa } \right)}}{{\sqrt {{{\left( {\alpha {{\left| g \right|}^2}\kappa  + \sigma _w^2} \right)}^2} - {{\left( {\alpha {\Re}\left( {{g^2}\tilde \kappa } \right)} \right)}^2}} }}}.
\end{equation}

\section{Translation within power budget}
\label{AppendB}
In this appendix we present the proof of Remark \ref{remark:powertranslation}. It is straightforward to prove that the translation $\V{v} = \M{A}\V{w}$ does not change the variance/power but only introduce asymmetry/improperness. Considering the transformation caused by the translation ${v}=  \sqrt{1+\zeta} w_{I} + i  \sqrt{1-\zeta} w_{Q} $, the power/variance is given by 
 \begin{equation}
\sigma_{v}^2 = ({1+\zeta}) \sigma_{w_{I}}^2 +   ({1-\zeta}) \sigma_{w_{Q}}^2. 
 \end{equation}
 Using the symmetric nature of r.v. $w$ i.e., $\sigma_{w_{I}}^2  = \sigma_{w_{Q}}^2$, it is clear that $\sigma_{v}^2 = \sigma_{w}^2$. On the other hand, the pseudo-variance can be calculated as 
  \begin{equation} \label{eqB2}  
\tilde{\sigma}_{v}^2 = ({1+\zeta}) \sigma_{w_{I}}^2 -   ({1-\zeta}) \sigma_{w_{Q}}^2  +i2 \sqrt{1-\zeta^2} {E\left\{ w_{I} w_{Q} \right \}}.
 \end{equation}
Again, the symmetry implies ${E\left\{ w_{I}w_{Q}\right\}} = 0$. Thus, the circularity coefficient can be derived from \eqref{eqB2} i.e., $ {|\tilde{\sigma}_{v}^2|}/{\sigma_{v}^2 } = \zeta $. 

The same concept can be extended to the symmetric discrete constellations with uniform prior probabilities. Considering the transformation caused by the translation $v_m =  \sqrt{1+\zeta} x_{mI} + i  \sqrt{1-\zeta} x_{mQ}$, the power of the transformed constellation is given by 
 \begin{equation}
 P = \frac{1}{M} \left( ({1+\zeta}) \sum\limits_{m = 1}^M  x_{mI}^2 +   ({1-\zeta}) \sum\limits_{m = 1}^M  x_{mQ}^2  \right).
 \end{equation}
 Using the symmetric property of the original discrete constellation $ \sum\limits_{m = 1}^M  x_{mI}^2 = \sum\limits_{m = 1}^M  x_{mQ}^2 $, it is clear that the power is preserved as $P = \frac{2}{M} \sum\limits_{m = 1}^M  x_{mI}^2$. Moreover, the non-zero pseudo-variance is given by 
  \begin{equation} \label{eqB3}
\tilde{P} = \zeta P + \frac{2i}{M}  \sqrt{1-\zeta^2} \sum\limits_{m = 1}^M x_{mI}\,x_{mQ}.
 \end{equation}
Again, the symmetry implies $\sum\limits_{m = 1}^M x_{mI}x_{mQ} = 0$. Thus, the circularity coefficient can be derived from \eqref{eqB3}, i.e., ${{|\tilde{P}|}/{P} = \zeta }$. 
\section{KKT Conditions}
\label{AppendC}
The convex non-linear constraint problem \textbf{P1a} can be efficiently solved using the first order necessary KKT conditions. We begin by writing the Lagrangian function $\mathcal{L}$ as
\begin{align} \label{C1}
\mathcal{L}& \left(   {\mathbf{p}},  {\lambda _1},{\lambda _2},{\lambda _3}\right) = \!  {\tilde{\rm{P}}_{\rm b}^{\rm UB}} \! \left(\! {\mathbf{p}},{\mathbf{p}}^{(k)}   \right)\! +\! {\lambda _1}\! \left(\! {\sum\limits_{m = 1}^M  {{\left| {{x_m}} \right|}^2}{p_m}\!  - \! 1\!} \right)\nonumber \\ 
&\qquad \quad +\!  {\lambda _2}\! \left( {\sum\limits_{m = 1}^M  {p_m}\!  - \! 1} \right)\! +\!  {\lambda _3} \! \left(  \log_2\left({M_{\rm u}}\right)\! -\! {\rm H}({\mathbf{p}}) \right),
\end{align} 
where the Lagrange multipliers are ${\lambda _1},{\lambda _2}$, ${\lambda _3} \geq 0$. 
Next. we evaluate the gradient of the \eqref{C1} with respect to the optimization variables in ${\mathbf{p}}$
\begin{equation}
 \nabla_{{\mathbf{p}}} {\mathcal{L}} = \left[  \frac{{\partial\mathcal{L}}}{{\partial {p_1}}} \quad  \frac{{\partial \mathcal{L} }}{{\partial {p_2}}} \quad \ldots \quad  \frac{{\partial \mathcal{L}}}{{\partial {p_{M_{\rm nu}}}}}  \right],
\end{equation}
where the partial derivative of ${\mathcal{L}}$ with respect to ${p_m}$ is given by
\begin{align}
\frac{{\partial \mathcal{L}}}{{\partial {p_m}}} =& \frac{{\partial {{\rm{P}}_{\rm b}^{\rm UB}}\left( {\mathbf{p}}^{(k)}   \right)}}{{\partial {p_m}}} + {\lambda _1}{\left| {{x_m}} \right|^2} +{\lambda _2} \nonumber  \\
& + {\lambda _3} \left( {\frac{1}{{\ln (2)}} \!+ \!{{\log }_2}\left( {{p_m}} \right)} \right), \quad \forall\, 1 \leq m
\leq M_{\rm nu}
\end{align}
Suppose that there is a local solution ${\mathbf{p}}^*$ of \textbf{P1a} and the objective function ${\tilde{\rm{P}}_{\rm b}^{\rm UB}} \! \left(\! {\mathbf{p}},{\mathbf{p}}^{(k)}   \right) $ along with the constraints \eqref{eq.pc} and \eqref{eq.rc} are continuously differentiable. Then, there exists a Lagrange multiplier vector $\lambda^*$, with components $\lambda_i$, where $i \in \left( 1,2,3 \right)$, such that the necessary first order KKT conditions (as presented in Table \ref{tab:TableKKT}) are satisfied at $\left(   {\mathbf{p}}^*,  {\lambda ^*}\right)$.
Interestingly, the KKT conditions are satisfied with 
\begin{align}
&\nabla_{\mathbf{p}}\mathcal{L} \left(   {\mathbf{p}}^*,  {\lambda _1^*},{\lambda _2^*,{\lambda _3^*}}\right)=0,  \label{C2}\\
& {\sum\limits_{m = 1}^M  {{\left| {{x_m}} \right|}^2}{p_m^*} } = 1,  \\
&   {\sum\limits_{m = 1}^M  {p_m^*} }  = 1, \\
& {\rm H}({\mathbf{p}}^*) = \log_2\left({M_{\rm u}}\right). \label{C3}
\end{align}
owing to the maximum transmission power preference, equality constraint and \ac{BER} -Rate trade-off , respectively. Thus, the $M_{\rm nu}+3$ solution parameters $\left(p_1^*,p_2^*,\ldots,p_{M_{\rm nu}}^*,  {\lambda _1^*},{\lambda _2^*,{\lambda _3^*}}\right)$ can be efficiently obtained by solving equations \eqref{C2}-\eqref{C3} using Levenberg-Marquardt algorithm \cite{marquardt1963algorithm}.

\section{Gradient for Optimization}
\label{AppendD}
The gradient of the upper bound on \ac{BER}  w.r.t \ac{GS} parameters is given as
\begin{align}\label{eq34}
 \nabla_{\mathcal{G}} {\rm{P}}_{\rm b}^{\rm UB} &= \left[  \frac{{\partial {\rm{P}}_{\rm b}^{\rm UB}}}{{\partial {\zeta}}} \quad  \frac{{\partial {\rm{P}}_{\rm b}^{\rm UB}}}{{\partial {\theta}}}   \right],\nonumber \\ & =
\frac{1}{\log_2 \left( M \right)}  \sum\limits_{m = 1}^{M} {\sum\limits_{n=1 \atop n \ne m}^{M}{ \Delta_{mn} }}  \left[ \frac{\partial {\gamma _{mn}}}{{\partial {\zeta}}} \quad  \frac{\partial {\gamma _{mn}}}{{\partial {\theta}}}   \right],
\end{align}
where  $\Delta_{mn}$ is the common part in both partial derivatives.
\begin{equation}
\Delta_{mn} = \frac{p_m \gamma_{mn}^{-3/2}}{2\sqrt{2\pi}} \sqrt{\frac{1-\rho_z^2}{\alpha}} e^{-\frac{\Omega_{mn}^2}{2}} \left( \ln \left( {\frac{{{p_m}}}{{{p_n}}}} \right)- \frac{1}{2{{\beta _{mn}^2}} } \right).
\end{equation}
Moreover, the partial derivative of $\gamma_{mn}$ with respect to the translation parameter $\zeta$ is given as
\begin{equation}
 \frac{\partial {\gamma _{mn}} }{\partial \zeta}= \frac{{{\bar{\xi}_{mnI}}^2}}{{\sigma _{\rm{I}}^2}}  + \frac{{2{\rho _z}{{\bar{\xi}_{mnI}}}{{\bar{\xi}_{mnQ}}}}}{{{\sigma _{\rm{I}}}{\sigma _{\rm{Q}}}}}  \frac{\zeta}{\sqrt{1-\zeta^2}} - \frac{{{\bar{\xi}_{mnQ}}^2}}{{\sigma _{\rm{Q}}^2}},
\end{equation}
where, ${{\bar{\xi}_{mnI}}} = {{\xi_{mn}}_I}\cos(\theta)-{{\xi_{mn}}_Q}\sin(\theta) $ and ${{\bar{\xi}_{mnQ}}}= {{\xi_{mn}}_I}\sin(\theta)+{{\xi_{mn}}_Q}\cos(\theta)$.
Furthermore,  the partial derivative of $\gamma_{mn}$ with respect to the rotation parameter is evaluated as in \eqref{eq:gammaderivatives} in the top of next page.

\begin{floatEq}
\begin{align}
 \frac{\partial {\gamma _{mn}} }{\partial \theta} = 2 \frac{1+\zeta}{{\sigma _{\rm{I}}^2}}\left(   {{\xi_{mn}}_I}\cos(\theta)-{{\xi_{mn}}_Q}\sin(\theta)     \right)\left(   -{{\xi_{mn}}_I}\sin(\theta)-{{\xi_{mn}}_Q}\cos(\theta)\right)+ \nonumber \\
+2 \frac{1-\zeta}{{\sigma _{\rm{Q}}^2}}\left(   {{\xi_{mn}}_I}\sin(\theta)+{{\xi_{mn}}_Q}\cos(\theta)     \right)\left(   {{\xi_{mn}}_I}\cos(\theta)-{{\xi_{mn}}_Q}\sin(\theta)\right)+ \nonumber \\
-\frac{{2{\rho _z}}}{{{\sigma _{\rm{I}}}{\sigma _{\rm{Q}}}}}  {\sqrt{1-\zeta^2}}
\left(   {{\xi_{mn}}_I^2}\cos(2\theta)-{{\xi_{mn}}_Q^2}\cos(2\theta)   -2{{\xi_{mn}}_I}{{\xi_{mn}}_Q} \sin(2\theta)  \right).
\label{eq:gammaderivatives}
\end{align}
\end{floatEq}

\bibliographystyle{IEEEtran}
\bibliography{IEEEabrv,References_PSvsGS_2020}

\end{document}